\documentstyle[12pt,epsf]{article}

\def\footnoterule{\kern-3pt\hrule\kern3pt}
\begin{document}
\def\bnu{\bar{\nu}}
\def\footnoterule{\kern-3pt\hrule\kern3pt}
\newfont\figfont{cmr7 scaled 1200}

\renewcommand{\thefootnote}{\fnsymbol{footnote}}

\rightline{MIT-CTP \#2257}
\rightline{hep-ph/9405217}
\rightline{April, 1994}

\vspace*{.3in}

\begin{center}

{\LARGE Using Heavy Quark Fragmentation into Heavy Hadrons}\\
\vspace{3pt}
{\LARGE to Determine QCD Parameters}\\
\vspace{3pt}
{\LARGE and Test Heavy Quark Symmetry}%
\footnote{This work is supported in part by funds provided by the U.S.
Department of Energy (DOE) under contract \#DE-AC02-76ER03069,
by the Texas National Research Laboratory Commission
under grant \#RGFY92C6, by the National Science Foundation
under Grant No. PHY89-04035 and by CICYT (Spain) under grant \#AEN 93-0234.
\hfil\break
\vskip 0.05cm
\noindent $^{\dagger}$National Science Foundation Young Investigator
Award.\hfil\break
Alfred P.~Sloan Foundation Research Fellowship.\hfill\break
Department of Energy Outstanding Junior Investigator Award.\hfil\break} \\
\vspace{20pt}
{\large L.~Randall$^{\dagger \ a,b}$  and N. Rius $^a$ \\
\vspace{3pt}
$^a$ Mass. Inst. of Technology, {\it  Cambridge MA 02139, USA} \\
$^b$ Institute for Theoretical Physics, University of California,
{\it  Santa Barbara \\ CA 93106, USA}} \\
\end{center}

\setcounter{page}{0}
\thispagestyle{empty}
\vspace{20pt}

\begin{abstract}
We present a detailed analysis of the use of heavy quark
fragmentation into heavy hadrons for testing the heavy quark effective
theory through comparison of the measured fragmentation
parameters of the $c$ and $b$ quarks. Our analysis is
entirely model independent. We interpret the known perturbative
evolution in a way useful for exploiting heavy quark symmetry at low
energy. We first show
consistency with perturbative
QCD scaling for measurements done
solely with $c$ quarks.
 We then apply the perturbative analysis and the heavy quark expansion
to relate measurements from ARGUS and LEP.  We
place bounds on a nonperturbative quark mass
suppressed parameter, and compare the values
for the $b$ and $c$ quarks.  We find consistency
with the heavy quark expansion but fairly sizable QCD uncertainties. We
also suggest that one might reduce the systematic uncertainty in
the result by not extrapolating to low $z$.
\end{abstract}

\vspace{\fill}
\begin{center}
{\it Submitted to:} Nuclear Physics B
\end{center}

\newpage
\renewcommand{\thefootnote}{\arabic{footnote}}
\setcounter{footnote}{0}

\newcommand{\beq}{\begin{equation}}
\newcommand{\eeq}{\end{equation}}
\def\bea{\begin{eqnarray}}
\def\eea{\end{eqnarray}}
\newcommand{\ba}{\begin{array}}
\newcommand{\ea}{\end{array}}
\newcommand{\ev}[1]{\langle #1 \rangle}
\def\sig{\sigma}
\def\sigi{\hat{\sigma}_i}
\def\siq{\hat{\sigma}_q}
\def\sn{\sigma_N}
\def\f{\hat f}
\def\dn{\hat{\Gamma}_N}
\def\am{\alpha(\mu)}
\def\a0{\alpha(\mu_0)}
\def\mev{{\rm MeV\ }}
\def\gev{{\rm GeV}}

\def\R{{\tilde R}}
\def\sd{M\frac{\partial}{\partial M}}
\def\M{\tilde{M}}

\section{Introduction}

Current $e^+ e^-$ colliders have excelled in the
precision study of weak interactions.
One can also hope to perform detailed tests of the predictions of QCD and
heavy quark symmetry.
With the copious production of heavy quarks,
the study of heavy quark fragmentation functions at
$e^+ e^-$ colliders might prove one such forum for  detailed
studies of QCD.

The factorization theorem states that
the measured fragmentation function can be written
as a convolution of two terms, one which depends on short-distance
physics and one which is sensitive to
large-distance dynamics.
The perturbative contribution has been understood for some time,
while heavy quark symmetry provides constraints on the nonperturbative
contribution.

The nonperturbative analysis that we follow was derived
in ref. \cite{jr}.
The basic idea is to evaluate the matrix elements of the heavy quark
operators at sufficiently low renormalization mass scales, where one
can exploit heavy quark symmetry to expand the moments
of the fragmentation function in $\Lambda/m$, where $\Lambda$ is some
QCD related scale, $m$ is the heavy quark mass, and the coefficients
in the expansion are mass independent.
With this parameterization of  the nonperturbative features of the
fragmentation function at low mass scale, one can  evolve the moments
via perturbative QCD to the high scale $Q^2$ at which
they are measured. One can then use the
experimental results
to extract the unknown nonperturbative parameters.
Since the same parameters enter the $c$ and $b$ quark fragmentation
functions and moments, one can test how well the heavy quark expansion
converges.

In this paper we apply this approach to the normalized
second moments of the $c$ and $b$ quark fragmentation functions,
measured at different center of mass energies.
We study the second moment because it is best measured
experimentally and because it involves a single
nonperturbative parameter at order $\Lambda/m$.

Our analysis consists of two parts. We first relate the
second moments of the $c$ quark fragmentation function measured
at different center of mass energies,
namely by ARGUS or CLEO and at PEP and LEP.
This part of the analysis involves solely perturbative QCD evolution.
In the second part of the analysis we combine both
perturbative QCD and the leading order
heavy quark expansion
to relate the moments of the $c$ and $b$ quark
fragmentation functions.

We will see that there is no inconsistency with preliminary
measurements. We will also see that by comparing the
fragmentation function of $c$ quarks at different center
of mass energy one might hope to constrain the QCD scale.
When using heavy quark symmetry to relate the $b$ and $c$
quark fragmentation functions, the calculation is less predictive,
due to the large uncertainty in higher order QCD effects at the
heavy quark mass scale. How useful the relations ultimately prove
is sensitive to the size of $\Lambda_{QCD}$.

In Section 2 we describe in detail the perturbative QCD scaling
accurate to subleading logarithmic order, and we summarize the
nonperturbative results of ref. \cite{jr} relevant for the purposes of
this paper.
In Section 3 we present our results.
Section 4 is a digression from our main analysis in which
we suggest that the experimental results be presented differently. Rather
than extrapolating the data to low $z$, we suggest a \lq \lq cutoff"
moment, where one only integrates from some cutoff value $z_0$.
We suggest that factorization, though no longer exact, might
work adequately.
We conclude in the final section.

\section{Theoretical Background}
\label{1}

In this section, we
give the theoretical framework with which we analyze heavy quark fragmentation.
We employ perturbative QCD in conjunction with the heavy quark
expansion to relate
the heavy quark fragmentation
functions measured at different center of mass energies.
Because it is best measured
and involves only a single nonperturbative parameter (if we only
keep the leading quark mass dependence), we will ultimately
relate not the fragmentation function itself, but the
normalized second moment.

The perturbative analysis
uses  refs. \cite{fpz}, \cite{mn} and the nonperturbative analysis
is from ref. \cite{jr}. From the latter reference, we use only
the results that the moments of the fragmentation
function have a heavy quark mass expansion, and
that the correct choice of fragmentation variable
corresponds to the ratio of energies. The
discussion which follows primarily summarizes the
perturbative scaling which was not done explicitly in ref. \cite{jr}.
We frame the analysis entirely in terms
of the standard procedure which is used when applying the heavy
quark theory, namely scaling, matching to the heavy quark theory,
and evaluating matrix elements in the heavy quark theory.
With the perturbative scaling, the measured fragmentation function  will
be used to determine the unknown nonperturbative parameters of
the fragmentation function. The tests of the theory come when
we relate the fragmentation function measured at different center
of mass energies, and when we relate the fragmentation parameters
for different flavors of heavy quark.

In this paper, we work to subleading
logarithmic order. We do not incorporate explicitly
Sudakov logarithms in the nonperturbative analysis,
since they are already included in the
parameterization of the heavy quark fragmentation
function at low renormalization mass scales.
Furthermore, the heavy quark expansion constrains us to
the study of low moments,
because of the large combinatoric factor which multiplies terms
in the expansion of higher order moments.
Therefore, as argued in ref. \cite{jr},
Sudakov logs are not important in our approach, since
low moments are not dominated by $z$ in the regime where Sudakov logs
are large.

We use the following notation:
$\sigma$ is the cross section for the inclusive
$e^+ e^-$ annihilation into a hadron, $H$,
$\sigi$ is the short-distance cross section for producing a parton $i$,
and $\f_i$ is the fragmentation function for producing the hadron
$H$ from the parton $i$.
According to the factorization theorem, the
cross section for the inclusive annihilation,
$e^+ e^- \rightarrow H + X$,
can be written as
\beq
{d \sigma(z,Q,m) \over dz} =
\sum_i \int_z^1
{dy \over y}
{d \sigi(y,Q,\mu) \over dy}
\f_i(z/y,\mu,m)
\label{e1}
\eeq
where $z$ is the ratio of the energy of the hadron to the beam energy
in the center of mass frame ($0 < z < 1$).
The factorization formula (\ref{e1}) states that the physical cross
section is the convolution of two terms.
The first, $d \sigi(y,Q,\mu) / dy$,
is the short-distance cross section for producing a parton $i$
(where $i$ can be a heavy or light quark or antiquark, or a gluon).
It is insensitive to the low-energy features of the process
and therefore does not depend upon the mass of the heavy quark.

The inclusive short-distance cross section for producing a parton
$i$ in $e^+ e^-$ annihilation via $\gamma$- and $Z$-exchange,
is given by
\beq
{d \sigi(z,\theta,Q) \over dz d\cos\theta} =
\lambda_i  \left\{
{3 \over 8} (1 + \cos^2\theta) \ C_{i,T}(z) +
{3 \over 4} \sin^2\theta \ C_{i,L}(z) +
{3 \over 4} \cos\theta \ C_{i,A}(z)
\right\}
\label{e2a}
\eeq
where $z=2q\cdot p/q^2$,
$p_\mu$ is the four-momentum of the parton,
$q_\mu$ is the four-momentum of the virtual boson,
with $\sqrt{q^2}=Q$,
and $\theta$ is the c.m. angle between the parton and the beam.
We have taken $\mu=Q$, so that
$d\siq(z,\theta,Q)$
is $\mu$--independent.
The notation $C_{i,1}(z) = C_{i,T}(z)$ and
$C_{i,2}(z) = C_{i,T}(z) - 2 C_{i,L}(z)$ is also used
in the literature.

In eq. (\ref{e2a}),
the factor $\lambda_q$ for quarks is just
the total cross section for producing a quark
\beq
\lambda_q \equiv
\hat{\sigma}_q =
{4 \pi \alpha_{em}^2 \over Q^2}
\{ e_e^2 e_q^2 +
2 e_e e_q v_e v_q {\rm Re} \chi(Q^2)
+ (v_e^2 + a_e^2) (v_q^2 + a_q^2) |\chi(Q^2)|^2 \}
\eeq
where
\beq
\chi(Q^2) = {Q^2 \over Q^2 - M_Z^2 + i Q^2 \Gamma_Z/M_Z}
\eeq
and $e_e$, $v_e$, $a_e$, $e_q$, $v_q$, $a_q$ are the charges and $Z$
couplings of the electron and quarks, respectively.
For gluons, $\lambda_G$ is given by
\beq
\lambda_G =  \sum_q
(\hat{\sigma}_q  + \hat{\sigma}_{\bar q} )
\eeq
Integrating (\ref{e2a}) over the angular variables we
obtain
\beq
{1 \over \lambda_i}  {d \sigi \over dz}
\equiv  C_i(z)
= C_{i,T}(z) + C_{i,L}(z)
\eeq
The coefficient functions $C_i$ are calculable in perturbative QCD as a
power series in $\alpha(\mu)$:
\beq
C_i(z,Q,\mu) =  C_i^{(0)}(z) +
{\alpha(\mu) \over 2\pi} C_i^{(1)}(z,Q,\mu) + \ldots
\label{e2}
\eeq
The leading order term in the quark coefficient function
$C_{q,L}(z)$ is suppressed by a factor
$m^2/Q^2$ and therefore negligible at high energies,
but the $\cal{O}(\alpha)$ contribution is not suppressed, so
at subleading order we need to include both contributions to $C_q$.
The coefficient function for gluons, $C_G$, is of order $\alpha$.

The second factor appearing in equation (\ref{e1}) is
the fragmentation function $\f_i(z,\mu,m)$, which
describes how the parton $i$
combines with surrounding partons to produce
the observed hadron H, in a much longer lapse of time.
It is insensitive to the high-energy part of the process and
therefore
depends on $m$ but not on   $Q$.

The scale $\mu$ is an arbitrary scale which separates the low- from the
high-energy dynamics.
If we know $\f_i(z,\mu_0,m)$ at some scale $\mu_0$, we can obtain
$\f_i(z,\mu,m)$ by using the Altarelli-Parisi
evolution equations, which organize
the large logarithms $\log(\mu/\mu_0)$
to the desired order in perturbation theory.
In practice, we will
work to subleading order, using the calculated anomalous dimensions
from ref. \cite{mn}.
As we will see, $C_i(z,Q,\mu)$ contains terms of the form
$\log Q^2/\mu^2$, so we will choose $\mu^2 \sim Q^2$ to avoid
large logarithms. On the other hand, we take
$\mu_0$ of order of the heavy quark mass, $m$,
where the results of the
heavy quark effective theory apply.

It is convenient to define
the singlet ($\f^S$) and non-singlet ($\f_q^{NS}$)
linear combinations of the
quark fragmentation functions:
\bea
\f_q^{(+)} &=& \f_q + \overline{\f}_q \nonumber  \\
\f^S  &=& \sum_{q=1}^{N_F}  \f_q^{(+)}  \\
\f_q^{NS} &=& \f_q^{(+)} - {1 \over 2 N_F} \f^S
\nonumber
\label{e3}
\eea
where $N_F$ is the number of flavours and
$\overline{\f}_q$ denotes the fragmentation function of the
antiquark $\bar{q}$.

According to the result of ref. \cite{jr}, we can expand
$\f_Q(z,\mu_0,m)$, the heavy quark fragmentation function,
for $\mu_0 \approx m$, \footnote{This expansion
is true in the HQET, so $\f_Q$ defined here is related
by a nontrivial perturbative matching condition to $\hat{f}_Q$
used in eq. (\ref{e1}).} as
\beq
\f_Q(z,m,m)={1\over \epsilon}
\hat{a}\left( {{1 \over z}-{m \over M_H} \over \epsilon}
\right) + \hat{b}\left( {{1\over z}-{m \over M_H} \over \epsilon}
\right)
\label{fragfcn}
\eeq
where $M_H$ is the mass of the heavy hadron and
$\epsilon \sim 1 - {m \over M_H}$.
The funny dependence on $z$ is because the expansion
is simplest in terms of the variable $1/z$.
It is important to use $z=E/E_{beam}$ (or its inverse) as the
fragmentation parameter. Other fragmentation variables
will permit nonvanishing higher twist contributions
at leading order in the heavy quark mass expansion and $\alpha_{QCD}$
\cite{jr}.
With this choice of fragmentation variable, higher twist
effects should be negligible (this is of course only significant
for relatively low $Q^2$
(relative to $m^2$) experiments, like ARGUS and CLEO).

We will not use the full form of this result, as the measured
fragmentation functions are too poor to do a detailed fit,
which involves many nonperturbative parameters. Instead,
as suggested in ref. \cite{jr}, we will consider only the  moments,
defined as
\beq
\dn^{HQ}(\mu,m)= \int_0^1 dz z^{N-1}
\f_Q(z,\mu,m)
\label{ea7}
\eeq
The low order moments have the advantage of being better measured, and
of being describable in terms of a small number of nonperturbative
parameters (when the heavy quark expansion is exploited).

According to the assumptions of the heavy quark theory, the
fragmentation functions(and their moments)
 for the light quarks and gluons at
the low renormalization scale can be taken to vanish.

By taking the moments of the function of eq. (\ref{fragfcn}), we derive
a heavy quark expansion for each of the moments. In general,
at order $\Lambda/m$, there are two independent parameters
in terms of which any moment can be expanded with known $N$ dependence
\cite{jr}.
However,
since we will  restrict our  numerical analysis to  the measured
second moment,
it is convenient to  define a single nonperturbative parameter
and expand the second moment as
\beq
\hat{\Gamma}_2^{HQ}(m,m)=1-{a \over m}
\label{defa}
\eeq
 From previous work \cite{heavymass}, one can also conclude that
there is a heavy quark mass expansion for the moments. The power
of the approach in \cite{jr} is that one can address the
higher twist corrections to determine the best choice of
fragmentation variable, that one can relate higher order
moments, and the one can in principle include higher order
mass suppressed terms.
The parameter $a$ is the single nonperturbative parameter
which is required to describe the second moment of the
fragmentation function if we work at order $\Lambda/m$.
Notice that because the fragmentation function is not defined in
terms of
an on shell matrix element,
the parameter $a$ is
not determined theoretically, but can be fit.
(This is in contrast with the distribution function,
for which one can derive $a(x)=\delta(x-m/M)$, with corrections
of order $(\Lambda/m)^2$.)
In ref. \cite{jr}, $a$ was defined in units of
$\overline{\Lambda}= M_H - m$.
In this paper, we have absorbed $\overline{\Lambda}$
in the definition of $a$, which is now a dimensional parameter.
It should
be kept in mind however, that $a$ is expected to be of order $\Lambda_{QCD}$.
The parameter
$a$ depends on the hadron type. Higher order terms can be incorporated--however
if the heavy quark expansion is valid, these terms should be suppressed,
and we should get an answer accurate at about the 10~\% level
by just keep the first mass suppressed contribution.

This expansion is valid only for $\mu_0\approx m$.
It is therefore necessary to match onto the full theory, and scale
the result to $\mu\approx Q$.  We now
describe the perturbative scaling and matching procedure.

The fragmentation functions
obey the Altarelli-Parisi evolution equations
\bea
{\partial \f^\tau (z,\mu,m) \over \partial \log \mu^2} &=&
\sum_{\tau'} \int_z^1 P_{\tau \tau'}(z/y,\am) \f^{\tau'} (y,\mu,m)
{dy \over y}
\hspace{1cm} \tau, \tau' = S,G
\label{e4}  \\
{\partial \f_q^{NS}(z,\mu,m) \over \partial \log \mu^2} &=&
\int_z^1 P_{NS}(z/y,\am) \f_q^{NS}(y,\mu,m)
{dy \over y}
\label{e5}
\eea
Only singlet quarks and gluons mix under evolution
\footnote{Note that the anomalous dimension matrix
for timelike processes is transposed aside from the dependence on
$N_f$.};
the non-singlet equation (\ref{e5}) decouples and can be solved
independently.

The  evolution functions
$P_{\tau \tau'}(z,\am)$ and $P_{NS}(z,\am)$
can be expanded in a power series in $\am$:
\beq
P(z,\am) = {\am \over 2 \pi} P^{(0)}(z) +
\left({\am \over 2 \pi} \right)^2 P^{(1)}(z) +
{\cal O} (\alpha^3)
\label{e6}
\eeq
The leading log contributions
($P_{\tau\tau'}^{(0)}$, $P_{NS}^{(0)} = P_{SS}^{(0)}$)
are the well-known Altarelli-Parisi splitting
functions \cite{ap}.

The definitions of $C_i$ and $\f_i$, and consequently of
$P(z,\am)$, are unique only at leading log order.
At the next-to-leading level they depend on the
factorization scheme used to separate the collinear singularities.
There are complete calculations available in the literature
in two different factorization schemes, which we will denote
I and II for definiteness.
Scheme I is defined in \cite{cfp} by
Curci, Furmanski and Petronzio, where they give the quark coefficient
functions and non-singlet evolution function
in $z$-space, as well as
their Mellin transforms to $N$-space.
The evolution function matrix in $z$-space for the singlet sector, including
quark and gluon mixing, has been computed in \cite{fp} and
the gluon coefficient functions can be found in \cite{fpz}.
Matching conditions for quarks and gluons within this scheme
are calculated in \cite{mn}. Alternatively, in \cite{frs}
Floratos, Ross and Sachrajda define Scheme II, which is used
in \cite{fkl} to compute coefficient functions for quarks and gluons
as well as non-singlet and singlet evolution functions both in $z$-space
and $N$-space
\footnote{However,
expressions for the timelike moments of the coefficient
functions and anomalous dimensions given in
the appendix B of ref. \cite{fkl} correspond to
$C_N^{(1)}$ and $P_N^{(1)}$, despite the $N+1$ index.}.

The results for $P_{\tau\tau'}^{(1)}$ in
\cite{fp} and \cite{fkl}
do not agree; there is a difference for
$P_{GG}^{(1)}(z)$ in the term proportional to the colour
factor $C_G^2$.
The timelike evolution functions (\ref{e6})
are obtained from the spacelike ones
(adequate to deep inelastic scattering processes)
by analytic continuation. The discrepancy mentioned above
is in fact a consequence of the discrepancy in the corresponding
spacelike evolution functions previously computed \cite{fpz}.

In principle one should evolve the fragmentation functions
$\f_i(z,\mu,m)$ by using eqs. (\ref{e4}) and (\ref{e5}),
i.e., including mixing between singlet quarks and gluons.
However we have explicitly
verified using Scheme I
that incorporating mixing never modifies our result
by more than about
1\% for the $b$ quark and 8\% for the $c$ quark.
Therefore, for the purpose of an expansion to order
$\Lambda/m$, we could safely neglect mixing.
The reason mixing is small is
that the gluon fragmentation function is never very big
and the gluon coefficient function is $\cal{O}(\alpha)$.
Our result agrees with the fact
that the probability of producing a heavy quark indirectly, via
a secondary gluon, is very small at $Q^2 \sim M_Z^2$.
The authors of ref. \cite{al} have shown that
the so-called Branco et al. terms, which contribute to
the heavy quark inclusive cross section,
can not be larger than 0.5\% for $b$-quarks and 4\% for $c$-quarks
and one would expect that the remaining terms are of the same
order of magnitude. This is very important
to our analysis, since it means that  we do not need to scale
the first moment (or equivalently the normalization factor) between
different momentum scales. If we are only dealing with energies less
than or equal to
the $Z$ mass, the normalization is approximately
independent of energy scale; that is very few  additional heavy
quarks are produced through QCD processes.
Therefore, we consider only  non-singlet evolution in the
remainder of this paper.

Beyond leading log level, the non-singlet evolution function has an
additional contribution due to $q \bar q$ mixing.
However, we have found that it is negligible
at the energies which are currently of interest.
Therefore, we use the non-singlet anomalous dimensions
given in \cite{mn}, which correspond to the factorization scheme I
and do not include $q \bar q$ mixing.

We are interested in the evolution of the moments of the
heavy quark fragmentation function, $\dn(\mu,m)$.
The next-to-leading order evolution equation for $\dn$ has a very simple
form:
\beq
{\partial \dn(\mu,m) \over \partial \log \mu^2} =
P_N(\am) \dn(\mu,m)
\label{e8}
\eeq
where
\beq
P_N(\am) = {\am \over 2 \pi} P_N^{(0)} +
\left({\am \over 2 \pi} \right)^2 P_N^{(1)}
\label{e9}
\eeq
is the Mellin transform of the Altarelli-Parisi splitting function in
next-to leading order.
The formula for $P_N^{(1)}$ is given in the appendix of ref. \cite{mn}.
The expression for $\am$ accurate to
next-to-leading order is
\beq
\am = {1 \over b_0 \log(\mu^2/\Lambda^2)}
\left( 1 - {b_1 \log[\log(\mu^2/\Lambda^2)] \over
b_0^2 \log(\mu^2/\Lambda^2)}  \right)
\label{e10}
\eeq
with
\beq
b_0 = {33 - 2 N_F \over 12 \pi}
\hspace{1.cm}
b_1 = {153 - 19 N_F \over 24 \pi^2}
\label{e11}
\eeq
Then eq. (\ref{e8}) takes the form
\beq
{\partial \dn(\mu,m) \over \partial \log \mu^2} =
\left[ P_N^{(0)} + {\am \over 2\pi}
\left( P_N^{(1)} - {2\pi b_1 \over b_0} P_N^{(0)} \right) \right]
\dn(\mu,m)
\label{e12}
\eeq
which can be solved analytically, yielding
\beq
\dn(\mu,m) =  \dn(\mu_0,m)
\left[ {\a0 \over \am} \right]^{{P_N^{(0)} \over 2\pi b_0}}
exp \left\{ {\a0 - \am \over 4\pi^2 b_0}
\left( P_N^{(1)} - {2\pi b_1 \over b_0} P_N^{(0)} \right) \right\}
\label{e13}
\eeq

The physical fragmentation function (eq. (\ref{e1}))
is obtained by convolution of the
fragmentation function with the hard cross-section, evaluated at the
scale $\mu$.
Therefore, the normalized moments
\beq
\sn(Q,m)= {1 \over \sigma}
\int_0^1 dz z^{N-1} {d\sig (z,Q,m) \over dz}
\label{e14}
\eeq
are given by
\beq
\sn(Q,m) =  C_N(Q,\mu)
\left[ {\a0 \over \am} \right]^{{P_N^{(0)} \over 2\pi b_0}}
exp \left\{ {\a0 - \am \over 4\pi^2 b_0}
\left( P_N^{(1)} - {2\pi b_1 \over b_0} P_N^{(0)} \right) \right\}
\dn(\mu_0,m)
\label{e15}
\eeq
where $C_N(Q,\mu)$ is the Mellin transform of the coefficient functions
(eq. (\ref{e2})),
\beq
C_N(Q,\mu) = 1 +
{\alpha(\mu) \over 2\pi} C_N^{(1)}(Q,\mu)
+ {\cal O} (\alpha^2)
\label{e16}
\eeq
The expression for $C_N^{(1)}(Q,\mu)$ is \cite{mn}
\footnote{$C_N^{(1)}(Q,\mu) =  \hat{a}_N^{(1)}(Q,\mu)$
in the notation of ref. \cite{mn}.}
\bea
C_N^{(1)}(Q,\mu) &=& C_F
\left[ \log {Q^2 \over \mu^2}
\left( {3 \over 2} + {1 \over N(N+1)} - 2 S_1(N) \right) \right.
\nonumber  \\
& & \left. + S_1^2(N) + S_1(N) \left({3 \over 2} - {1 \over N(N+1)} \right)
+ 5 S_2(N) \right. \nonumber  \\
& & \left. + {1 \over N} - 2 {2N+1 \over N^2(N+1)^2}
+ {1 \over (N+1)^2} -  {3 \over 2} {1 \over N+1} -{9 \over 2}
\right]
\label{e17a}
\eea
where $C_F=4/3$ and
\beq
S_l(N) = \sum_{j=1}^N {1 \over j^l}
\eeq
We will take  $\mu=Q$ from now on,
so that  $C_N^{(1)}$ does not depend upon $\mu$.
We therefore  use the notation
$C_N^{(1)} = C_N^{(1)}(\mu,\mu)$.

In eq. (\ref{e15}), $\dn(\mu_0,m)$ is the moment of the
heavy quark fragmentation function in the full theory at the scale
$\mu_0 \sim m$.
Note that it is critical to the use of a subleading log calculation that
we know the $\mu$--dependence of the matrix element.  Even though
the matrix element itself is nonperturbative, the scale dependence
is not, and is contained in the matching coefficient. We note that
in reality, there is still some scaling which goes on below the
heavy quark mass scale because of the scale dependence
of the spin--dependent mass suppressed operator \footnote{We thank
Eric Braaten for pointing this out.}. However,
since the maximum scaling between the $b$ and $c$ mass scales
 of the gluon magnetic moment
operator is 20\%, and it is only
one of several contributions to the $\Lambda/m$ term, this
effect is never numerically more significant than the higher
order mass suppressed corrections.
 We therefore neglect scaling in the heavy quark theory itself;
the $\mu$ dependence is  then just that from the matching between the full
and heavy quark theories.
Because this matching is done at a fairly high momentum scale, of
order the heavy quark mass, the scale dependence is calculable order
by order in perturbation theory.
We then have
\beq
\dn(\mu_0,m) = \left(1 +
{\a0 \over 2 \pi} d_N^{(1)}(\mu_0, m) \right)
\dn^{HQ}(m)
\label{e17}
\eeq
We take $\dn^{HQ}(m)$ as independent of $\mu_0$, but because we
work at a finite order in perturbative theory we will see that
there is a fairly substantial dependence
on $\mu_0$ which cannot be neglected.
This will be clear in section~\ref{2}.

The matching condition which agrees with the factorization
prescription for $C_N^{(1)}(Q,\mu)$ and $P_N^{(1)}$
can be obtained from ref. \cite{mn} and reads
\bea
d_N^{(1)}(\mu_0,m)
&= &\log{\mu_0^2 \over m^2} P_N^{(0)}
+ C_F \left[ \left. -2 S_1^2(N) + 2 S_1(N) \left( 1 + {1 \over N(N+1)} \right)
\right.\right. \nonumber  \\
& & \left. - 2 S_2(N) - {2 \over (N+1)^2} - {1 \over N(N+1)} + 2 \right]
\label{e18}
\eea
where
\beq
P_N^{(0)} = C_F
\left( {3 \over 2} + {1 \over N(N+1)} - 2 S_1(N) \right)
\eeq
Using (\ref{e17}) we can write eq. (\ref{e15}) as
\beq
\sn(Q,m) \equiv \ev{z^{N-1}(Q,m)} =
\ev{z(Q,m)^{N-1}}_{pert}\ev{z(m)^{N-1}}_{nonpert}
\label{endresult}
\eeq
where
\bea
\ev{z(Q,m)^{N-1}}_{pert} &=&  C_N(Q,\mu)
\left[ {\a0 \over \alpha(Q)} \right]^{{P_N^{(0)} \over 2\pi b_0}}
\left( 1 + {\a0 \over 2 \pi} d_N^{(1)}(\mu_0, m) \right)
\nonumber \\
&\times& exp \left\{ {\a0 - \alpha(Q) \over 4\pi^2 b_0}
\left( P_N^{(1)} - {2\pi b_1 \over b_0} P_N^{(0)} \right) \right\}
\label{er1}
\\
\ev{z(m)^{N-1}}_{nonpert}&=&\dn^{HQ}(m)
\label{er2}
\eea
It is important to clarify the issue of the factorization
scheme dependence.
The separate results for the anomalous dimensions and the
coefficient functions depend on the factorization scheme,
but in the convolution there is a cancellation of the scheme dependence
order-by-order in $\alpha$ \cite{fpz}.
Neglecting next-to-next-to-leading terms and using eq. (\ref{e16}),
we can write the coefficient of
$\alpha(Q)$ in the exponential of eq. (\ref{er1}) as
$P_N^{(1)} - 2\pi b_0  C_N^{(1)}$.
This combination can be shown to be independent of the renormalization
scheme at the subleading level,
and we have explicitly checked that by using both schemes I and II.
However the coefficient of $\a0$ still
depends on the factorization scheme and
therefore so does the matching condition (\ref{e18}).
In order to obtain an expression in terms of
renormalization scheme independent quantities, we use the equation
\beq
C_N^{HQ}(\mu_0,m) = 1 +
{\a0 \over 2 \pi} d_N^{(1)}(\mu_0,m) +
{\a0 \over 2 \pi} C_N^{(1)}
\label{e19}
\eeq
where $C_N^{HQ}(\mu_0,m)$ is the short-distance cross section
for producing a heavy quark of mass $m$.
We then rewrite eq. (\ref{er1}) as
\bea
\ev{z(Q,m)^{N-1}}_{pert} &=&  C_N^{HQ}(\mu_0,m)
\left[ {\a0 \over \alpha(Q)} \right]^{{P_N^{(0)} \over 2\pi b_0}}
\nonumber \\
&\times& exp \left\{ {\a0 - \alpha(Q) \over 4\pi^2 b_0}
\left( P_N^{(1)} - 2\pi b_0 C_N^{(1)} -
{2\pi b_1 \over b_0} P_N^{(0)} \right) \right\}
\label{sir}
\eea
Eqs. (\ref{sir}) and (\ref{er1}) are equivalent at the subleading
level. Moreover, the evolution factor entering in eq. (\ref{sir})
is now independent of the factorization scheme and thus the matching
condition (\ref{e19}), derived in the scheme of \cite{cfp},
is also adequate for the scheme used in \cite{fkl}.

However, since we are using results calculated within a
single scheme it is not necessary to use
scheme independent expressions as in (\ref{sir}).
In fact, this turns out to be the preferred procedure.
This is because in order to obtain a scheme independent result,
we added terms which are technically higher order, but
are in fact quite large, due to the large size of
$C_N^{(1)}$.
For example, ${\alpha(m_c/2) \over 2 \pi} C_N^{(1)}$
is as big as 0.8 for $\Lambda_5=$ 225 MeV.
So we are probably introducing spurious $\mu$ dependence by adding
these unduly large terms. We therefore used eq. (\ref{er1}) in our
calculations.

To summarize, we have shown that the measured moment of the
fragmentation function can be expressed as in eq. (\ref{endresult}).
The perturbative contribution $\ev{z(Q,m)}_{pert}$ is
given by (\ref{er1}) or (\ref{sir}),
where the former has smaller $\mu$ dependence and is
actually what one obtains on integrating the renormalization
group equation. Once
one has calculated $\ev{z(Q,m)}_{pert}$, one can extract
the nonperturbative parameter $a$, defined in eq. (\ref{defa}).
This parameter is independent of $Q$, the energy of the
experiment, and so long as one always deals with the same kind of hadron,
it should be also independent of $m$.

Unfortunately however there are still  fairly large QCD uncertainties
in the perturbative scaling given that we are working only to subleading
logarithmic order and that the QCD scale is not so well known.
The problem is that we
need to know $\ev{z(Q,m)}_{pert}$ sufficiently
well to extract $\ev{z(m)}_{nonpert}$
at the level of
$a/m$.  For large $m$, the required accuracy is
greater, though the perturbation theory is better.
We find that the error in extracting $a$ from the $b$ and
$c$ fragmentation functions is in fact comparable.
In order to determine the accuracy of our procedure, we
will consider different values of $\mu_0$.
Higher order perturbative effects may also be estimated
from the ${\cal O}(\alpha^2)$ terms neglected in eq. (\ref{sir}).
We find that these uncertainties are never as large as the $\mu_0$
dependence. So we estimate the uncertainty
from higher order effects by considering different
values for the  renormalization scale $\mu_0$.
The particular choices and results  are given in the next section.

\section{Results}
\label{2}

In this section, we apply the procedure described in the previous
section, where it was shown how
perturbative QCD and the heavy quark
expansion can be used to relate moments of
fragmentation functions of different flavors
measured at different $Q^2$ values.
These predictions can be compared with the data from ARGUS,
CLEO, PEP, PETRA and LEP.
Since current results from LEP  are only preliminary,
we present our analysis in such a way that
it can be readily applied with improved measurements. To
do this we center the values on our plots on the preliminary
measurement and we extract the nonperturbative parameter
over the full range of experimentally allowed numbers (that
is, the measured value with 2 sigma errors).

Recall the uncertainties in our predictions arise from two different
types of QCD uncertainties. There is uncertainty in the
perturbative part of the calculation  due to both
the poorly known value of $\alpha_{QCD}$ and the fact that
we work only to subleading logarithmic order. For quantities
involving perturbative scaling only for large $Q^2$ these
uncertainties are expected to be small.
But in order to apply heavy quark symmetries, one
always needs to scale to the quark mass scales. For such low
energy, higher order QCD corrections and the uncertainty in the
exact value of $\alpha_{QCD}$ can be important.

It therefore makes sense to divide our analysis into two parts.
We first consider simply measuring $\ev{z_c}$ at two different energy scales
(sufficiently larger than the $c$ quark mass that higher twist contributions
should be small). These values should be related by
 {\it only} perturbative QCD,
at scales for which the subleading calculation
 should prove reliable.
 The measurements  seem to
be consistent with perturbative QCD scaling
over the allowed range of $\Lambda_{QCD}$. Turned
around, it means that with sufficiently accurate data, one can constrain
$\Lambda_{QCD}$ (although probably not as well as with other methods).

Following this analysis, we proceed to incorporate the full formalism,
employing both perturbative QCD and a nonperturbative expansion
to relate $b$ and $c$ quark fragmentation functions.   Here we
will find sizable QCD uncertainty. Nonetheless, we will see
that the data is substantially self consistent.

Notice that throughout the analysis of section~\ref{1} we have assumed
fragmentation into a specific final state, although experiments at
very high center of mass energy such as LEP do not distinguish
hadron species.
This should not be a problem when comparing inclusive measurements
such as those done at LEP, since
the normalized second
moment for the inclusive measurement has the same heavy
quark expansion for the $b$ and $c$ quark.

This is potentially a problem if different
experiments select differently on the final hadronic state.
The problem occurs because the nonperturbative parameter $a$
depends on the hadron type. The nonperturbative contribution to
the mean $\ev{z}$ experimentally measured
can be a different linear combination of the different $a's$
at the  different experiments.
However, measurements from ARGUS
show \footnote{We thank G. Bonvicini for analyzing the
data from the Ph.D. Thesis of P. C-Ho Kim
and H.C.J. Seywerd (ARGUS Collaboration).}
that $\ev{z_D} \approx \ev{z_{D^*}}\approx
\ev{z_{\Lambda_c}}$ within errors. Preliminary measurements
from CLEO seem to also support this claim.  If
this is the case, we can to a good approximation neglect
the differences between the different hadron types
when comparing the measurements at different center
of mass energy.

Moreover, for inclusive measurements at sufficiently high
energy,
the dependence on hadron type can be ignored, since
the perturbative QCD scaling is the same for all hadrons
so that it is always the same linear combination
of $a's$ which is measured at any energy.
This is obvious, since $\mu$ can
be chosen so that only the coefficient function
depends on $Q$.

Finally, for measurements which select on a specific final state
(eg $D^*$), there is no problem.

Before we begin, we outline the differences in our approach from
previous studies \cite{cn}.
First, we make no attempt to fit the entire
curve. We look only at the mean $\ev{z}$, which is much
better measured than the fragmentation function itself. Furthermore,
the mean $\ev{z}$ depends only on QCD parameters and one
single nonperturbative parameter $a$, where $a$ is defined by
\beq
\ev{z(m)}_{nonpert} = 1-{a \over m}
\label{znp}
\eeq
Our result is therefore  independent of any assumed functional
form of the fragmentation function.  Furthermore, $\ev{z}_{nonpert}$
is much more stable against variations in $\mu$ and the QCD scale
than the parameters used to fit a particular functional form.
For example, in ref. \cite{cn}, where the functional form
$z_{nonpert}=z^\alpha (1-z)^\beta$ was used, the parameters $\alpha$
and $\beta$ varied enormously in comparison with
$\ev{z}_{nonpert}$.

Second, we use eq. (\ref{er1}) rather than (\ref{sir})
to calculate the perturbative factor $\ev{z(Q,m)}_{pert}$.
As explained earlier, this result is better behaved with respect to
$\mu$-variation, and it is in fact what is obtained
by integrating the renormalization group equations.

Third, we do not incorporate explicitly Sudakov logs in the
nonperturbative fragmentation function.

Fourth, we use the experimental results on the mean
$\ev{z} = \ev{E} / E_{beam}$.
It is important to use the variable $\ev{z}$ because its
non-perturbative contribution scale linearly in the mass of
the heavy quark, according to eq. (\ref{znp}).

Finally, we use updated experimental data from LEP which leads to a
harder $b$ quark fragmentation function. As a result we find better
agreement with the heavy quark effective theory predictions than in
ref. \cite{cn}.

It should be noted that there is some dependence in the measured $\ev{z}$
value on the assumed functional form (contributing to the systematic error)
because the data is not measured to arbitrarily low $z$, so
functional forms of the fragmentation function are assumed
in order to extrapolate to low $z$.  In
section~\ref{3}
we will show that one can use moments with a cutoff greater than
zero to reduce this dependence on the functional form.

In both parts of the following
analysis, we will allow for different values
of $\Lambda_{QCD}$ within the allowed range. We estimate
the importance of the neglected higher order perturbative QCD
contributions by testing the stability of our results with respect
to a change in renormalization scale. It is an important feature
of the analysis that the $\mu_0$ dependence is entirely perturbative
for a sufficiently heavy quark. Nonetheless, $\alpha(\mu_0)$
at the scale of the heavy quark mass is so large that it is not clear
how well the perturbation theory converges.
We estimate this uncertainty by allowing for a sizable
variation in $\mu_0$, between $2m$ and $m/2$.
This might be too large a range, but without a higher order
calculation, it is impossible to determine the accuracy of the
subleading calculation.

When we invoke the heavy quark expansion, we work in the
approximation of only retaining the leading mass dependent correction,
suppressed by a single power of the heavy quark mass.  Of course one can
readily incorporate higher order terms. But we would like to test whether
the heavy quark expansion works sufficiently well to get agreement (at
least at the 10~\% level) even without incorporating higher order
contributions.
This is important, as the heavy quark effective theory is still essentially
untested at the $\Lambda/m$ level.

\subsection{Perturbative Analysis: Relating $\ev{z_c}$ at Different
Energy Scales}
In this section we focus on what can
be learned using  solely  perturbative QCD
evolution, without the heavy quark effective theory.
Even if it turns out that QCD perturbation theory
at subleading order is not sufficiently accurate
at the low scale
$\mu_0^2 \sim m^2$, it certainly should give a good description
of the evolution between scales above the mass of the
$\Upsilon$.
Thus if we  compare
the moments of the $c$ quark fragmentation function measured
by CLEO or ARGUS, and at PEP, PETRA and LEP,
the measurements  should be  related through
the perturbative evolution between the scales of the experiments.
Again, this is with the caveat that if a different linear combination
of hadron states is selected, the relations only need hold
if the nonperturbative fragmentation function is the same for
each hadron type, as seems to be the case. Furthermore, since
$D^*$'s are the dominant contribution at these experiments, one
would expect the perturbative relation to hold fairly well.
In fact, with an exclusive measurement on $D^*$'s at both
experiments, the perturbative scaling result would certainly apply.

According to the evolution equations,
the $N=2$ measured moments of the fragmentation function for the
$c$ quark are given by
\beq
\ev{z_c(Q)} =  C_2(Q,\mu) \hat{\Gamma}_2(\mu,m_c)
\label{pt1}
\eeq
Therefore the ratio between the mean $\ev{z_c(Q)}$ measured at
different energy scales is
\beq
{\ev{z_c(Q_2)} \over \ev{z_c(Q_1)}}  =
{C_2(Q_2,\mu) \over C_2(Q_1,\mu)}
\label{pt2}
\eeq
We relate the $N=2$ moments measured at
ARGUS ($Q_1=10.6$ GeV) or CLEO ($Q_1=10.55$ GeV)
to those measured at PEP ($Q_2=29$ GeV) and LEP ($Q_2=91$ GeV).
Since the experimental data from ARGUS and LEP have smaller errors,
we focus our discussion mainly on the results from those experiments.
The difference between the perturbative calculation
for ARGUS and CLEO energies is negligible, so we only present the
theoretical results for $Q_1=10.55$ GeV.

Since for the theoretical computation of the
right hand side in eq. (\ref{pt2})
we only need to evolve the coefficient functions
between $Q_1^2$ and $Q_2^2$, which are well above $\Lambda_{QCD}$,
we expect the perturbative evolution with next-to-leading
accuracy to work well. The evolution equations
for the coefficient functions are related to that for the fragmentation
function by
\beq
{\mu d C_N(Q,\mu) \over d \mu}=-{\mu d \dn(\mu,m) \over d \mu}
\label{cf}
\eeq

We choose the scale $\mu \sim Q_1$, so
$C_2(Q_1,\mu)$ is given by eqs. (\ref{e16}) and (\ref{e17a}),
with $Q=Q_1$.
For $C_2(Q_2,\mu)$ we also use the evolution equation
(\ref{e15}) and eq. (\ref{cf}) to obtain
\beq
C_2(Q_2,\mu) = \left(1 + {\alpha(Q_2) \over 2\pi} C_2^{(1)} \right)
\left[ {\am \over \alpha(Q_2)} \right]^{{P_N^{(0)} \over 2\pi b_0}}
exp \left\{ {\am - \alpha(Q_2) \over 4\pi^2 b_0}
\left( P_N^{(1)} - {2\pi b_1 \over b_0} P_N^{(0)} \right) \right\}
\label{pt3}
\eeq
To account for theoretical uncertainties of the prediction,
we computed the ratio
$C_2(Q_2,\mu) / C_2(Q_1,\mu)$
using  the
values of $\Lambda$ in the five flavor theory,
$\Lambda_5$ = 75, 125, 175, 225, 275 MeV and,
as an estimate of higher order effects, we varied
the scale $\mu$ between $Q_1/2$ and $2 Q_1$.

The QCD perturbative results for the ratio
$C_2(Q_2,\mu) / C_2(Q_1,\mu)$, with $Q_1=10.55$ GeV and $Q_2=91$ GeV
are shown in Table 1. Notice that
the results do not depend strongly on the scale $\mu$,
as one would expect since subsubleading effects
at the energy scales we are considering should be rather small.

\begin{table}
\begin{center}
\begin{tabular}{||c|c|c|c||}     \hline
$\Lambda_5$ (MeV) & $\mu=Q_1/2$ & $\mu=Q_1$ & $\mu=2 Q_1$  \\ \hline
75  & 0.813 &  0.825 &  0.832   \\  \hline
125 & 0.794 &  0.808 &  0.816   \\  \hline
175 & 0.779 &  0.795 &  0.804   \\  \hline
225 & 0.766 &  0.784 &  0.794   \\  \hline
275 & 0.755 &  0.774 &  0.785   \\  \hline
\end{tabular}
\end{center}
\caption{$\displaystyle
{{C_2(Q_2=91 \gev,\mu) \over C_2(Q_1=10.55 \gev,\mu)}}$}
\end{table}
The mean $\ev{z_{D^{*+}}}$ for the fragmentation function
of a $c$ quark
into $D^{*+}$ mesons has been measured both at
ARGUS \cite{argus} and LEP \cite{aleph}.
The results are
\bea
\ev{z_{D^{*+}}(Q_1=10.6 \gev)}  &=& 0.647 \pm 0.006  \nonumber \\
\ev{z_{D^{*+}}(Q_2=91 \gev)} &=& 0.495 \pm 0.013
\label{ds}
\eea

\begin{figure}[ht]
\epsfxsize=3.6in
\epsfysize=3in
\hbox to \hsize{\hss\epsffile[24 68 574 738]{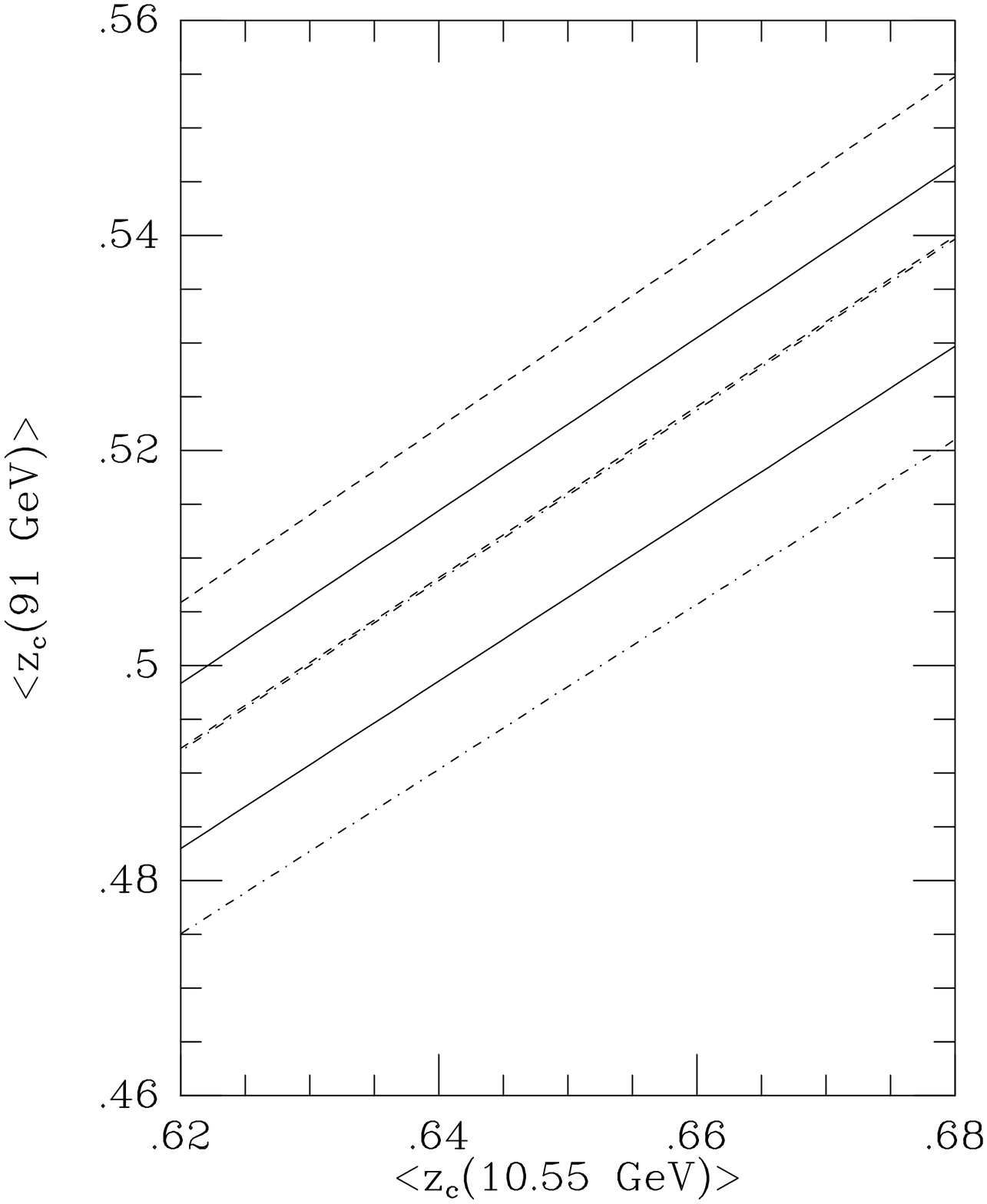}\hskip.5in\hss}
\vspace*{5mm}
\advance\leftskip by 0.36in
\advance\rightskip by 0.36in
{\figfont Figure 1: $\ev{z_c(Q_2=91 \gev)}$ as a function of $\ev {z_c
(Q_1=10.55 \gev)}$, for $\Lambda_5=$ 125 MeV (dashed line), 175 MeV (solid),
and 225 MeV (dashed-dotted). For each value of $\Lambda_5$ the lower line
corresponds to $\mu=Q_1/2$ and the upper line to $\mu=2 Q_1$.}
\end{figure}

In order to relate the inclusive measurements from PEP and LEP
with the ARGUS measurements of $\ev{z_D}$, $\ev{z_{D^*}}$
and $\ev{z_{\Lambda_c}}$
we have to average over all the final states.
\footnote{In fact, there are ARGUS measurements
only for $D^{*+}$ mesons, but we assume
$\ev{z_{D^{*0}}} = \ev{z_{D^{*+}}}$.}
If we assume heavy quark symmetry
and that a $c$ quark fragments into
baryons about 12\% of the time, the mean value of $z_c$ is
\bea
\ev{z_c(Q_1=10.6 \gev)} &=& 0.12 \ev{z_{\Lambda_c}}
+ 0.88
\left[ {1 \over 4} \ev{z_D} + {3 \over 4} \ev{z_{D^*}} \right]
\nonumber \\
&=&0.640  \pm  0.009
\eea
It would be very useful to test the assumed ratio of 3:1 for
$D^*$ relative to $D$ production is in fact correct.
Preliminary results from CLEO seem to give a bigger value,
so we allow for a wide
range $0.62 \leq \ev{z_c(Q_1)} \leq 0.68$.
The mean $\ev{z_c}$ measured inclusively at PEP \cite{pp} is
\beq
\ev{z_c(Q_2=29 \gev)} = 0.526 \pm 0.03
\eeq
while at LEP
\beq
\ev{z_c(Q_2=91 \gev)} = 0.487 \pm 0.011
\eeq
The numbers shown in Table 1 are all that is needed
to compare the measurements from ARGUS or CLEO and LEP.
In Figure 1 we present this result graphically  for
a slightly narrower range of $\Lambda_5$,
namely $\Lambda_5 = (175 \pm 50)$ MeV,
according to ref. \cite{pdb}.
For each value of
$\ev{z_c(Q_1=10.55 \gev)}$ given
on the horizontal axis, we plot the value of
$\ev{z_c(Q_2=91 \gev)}$ which should be measured according to
eq. (\ref{pt2}).
There are two lines for each value of $\Lambda_5$, corresponding
to $\mu=Q_1/2$ and $\mu=2 Q_1$.
So for a given value of $\alpha_{QCD}$
the prediction for $\ev{z_c(Q_2=91 \gev)}$
lies between these lines.

We see that from
the ARGUS measurement $\ev{z_c(Q_1=10.6 \gev)}$ =0.640,
$\ev{z_c(Q_2=91 \gev)}$ is predicted to be in the range
$0.490 \leq \ev{z_c(Q_2=91 \gev)} \leq 0.522$.
If we allow $\ev{z_c(Q_1=10.6 \gev)}$ to vary
within  $2 \sigma$ around the measured value
we obtain
$0.477 \leq \ev{z_c(Q_2=91 \gev)} \leq 0.537$.
Therefore, the results are consistent at the
$2\sigma$ level but they seem to favour large values of
$\Lambda_{QCD}$.

If we restrict the analysis to $D^{*+}$ mesons, from
$\ev{z_{D^{*+}}(Q_1=10.6 \gev)}$ = 0.647
we predict
$0.496 \leq \ev{z_{D^{*+}}(Q_2=91 \gev)} \leq 0.528$
($0.487 \leq \ev{z_{D^{*+}}(Q_2=91 \gev)} \leq 0.538$
within  $2 \sigma$).
It is very important to relate the exclusive measurements
on $D^{*+}$ mesons at both experiments, because in this case the
nonperturbative fragmentation function is exactly the same
(as it does not depend on $Q^2$) and we are sure to be testing
perturbative QCD, without invoking heavy quark symmetry
at all. With sufficient statistics, this could prove to
be a useful measurement of $\Lambda_{QCD}$.

\begin{figure}[ht]
\epsfxsize=3.6in
\epsfysize=3in
\hbox to \hsize{\hss\epsffile[24 68 574 738]{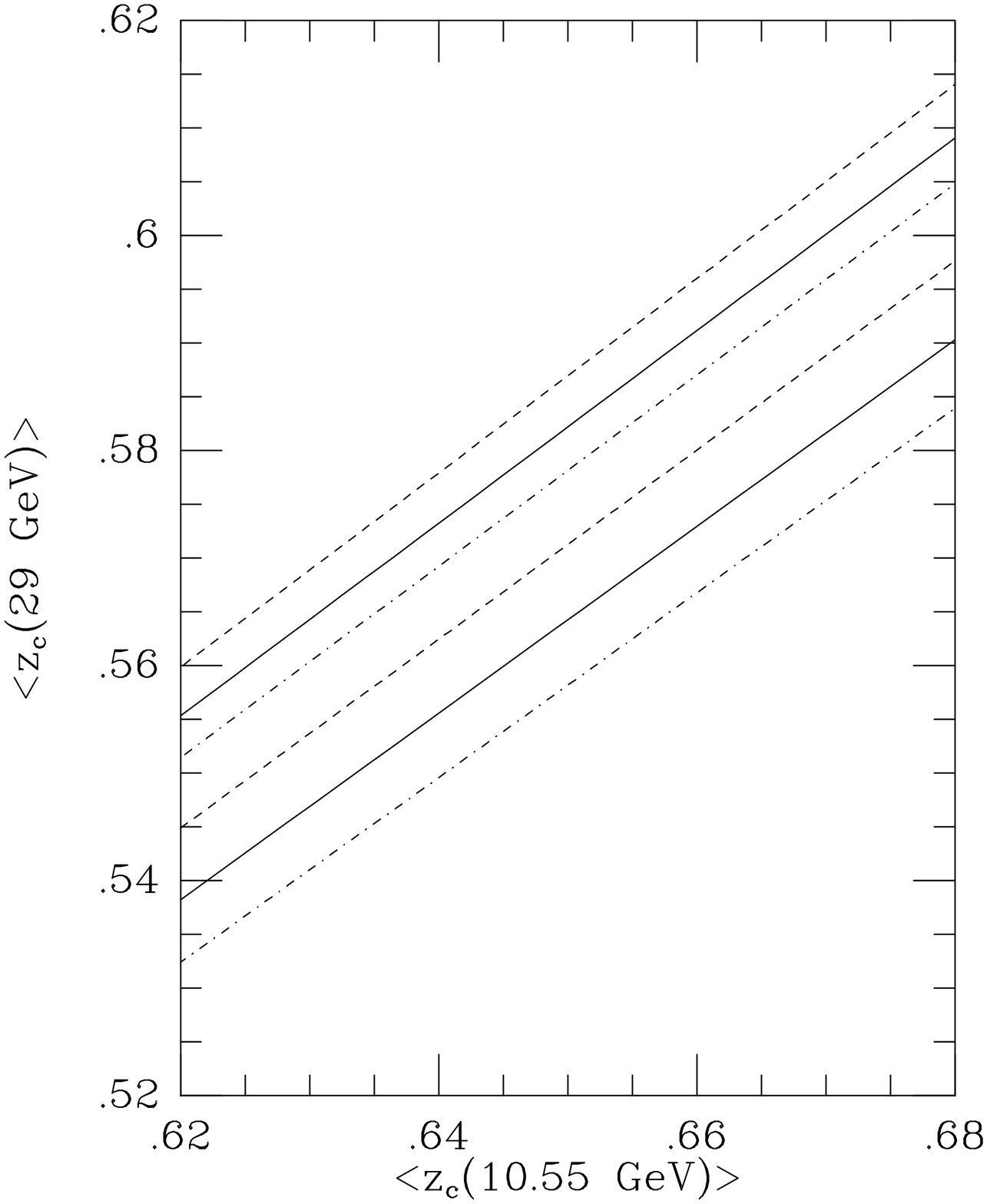}\hskip.5in\hss}
\vspace*{5mm}
\advance\leftskip by 0.36in
\advance\rightskip by 0.36in
{\figfont  Figure 2:
$\ev{z_c(Q_2=29 \gev)}$ as a function of $\ev{z_c(Q_1=10.55 \gev)}$, for
the same values of $\Lambda_5$ and $\mu$ as in Figure~1.}

\end{figure}

We have also related $\ev{z_c(Q_1=10.6 \gev)}$ and
$\ev{z_c(Q_2=29 \gev)}$. The result is shown in Figure~2,
for the same values of $\Lambda_5$ and $\mu$ as in Figure~1.
In this case we obtain
$0.549 \leq \ev{z_c(Q_2=29 \gev)} \leq 0.578$
from the ARGUS measurement $\ev{z_c(Q_1=10.6 \gev)}$ =0.640
($0.534 \leq \ev{z_c(Q_2=29 \gev)} \leq 0.594$
within  $2\sigma$).
We also see that although the QCD perturbative calculation
tends to give a too large prediction for
$\ev{z_c(Q_2=29 \gev)}$, since the experimental error of this
measurement is bigger than at LEP, the theoretical prediction lies
always within $2\sigma$ of the measured value.
Therefore, it is not possible to constrain further
$\Lambda_{QCD}$ by relating
$\ev{z_c(Q_1=10.55 \gev)}$ and $\ev{z_c(Q_2=29 \gev)}$.
Obviously, the same applies when we compare
$\ev{z_c(Q_1=29 \gev)}$ and $\ev{z_c(Q_2=91 \gev)}$.

One can do a similar analysis for the $b$ quark, relating
$\ev{z_b(Q_1=29 \gev)}$ measured at PEP and $\ev{z_b(Q_2=91 \gev)}$
from LEP. However, the QCD perturbative evolution is
in fact inconsistent with the experimental measurements
\cite{pp}, \cite{aleph}, namely
\bea
\ev{z_b(29 \gev)}&=&  0.715 \pm 0.03   \\
\ev{z_b(91 \gev)}&=&  0.714 \pm 0.012
\eea
Presumably, the $b$ measurement at 29 GeV is not reliable.

We conclude that
perturbative QCD predictions are consistent with current
data on $c$ quark fragmentation function within experimental errors.
However the data seems to favor large
$\Lambda_{QCD}$.
Figure~1 shows that a measurement of
$\ev{z_c}$ from CLEO would be very useful
in constraining further $\Lambda_{QCD}$.

The measured second moment
of the $b$ quark fragmentation function at $Q=29 \gev$
seem to be inconsistent with perturbative QCD and the
LEP measurement at $Q=91 \gev$.
For the $c$ quark, the second moment measured at $Q=29 \gev$
does not provide further bounds, since the
experimental error is larger than in later experiments.
Therefore, in the following section we only consider
the more accurate measurements
of the heavy quark fragmentation functions at center of mass energies
$Q=10.55$ GeV and $Q=91$ GeV.

\subsection{Relating $\ev{z_b}$ and $\ev{z_c}$}

In this section, we incorporate QCD directions and the leading
order heavy quark expansion in order to relate the measured
second moment of the $b$ and $c$ quark
fragmentation functions,
the first measured at LEP and
the latter measured at ARGUS, CLEO and LEP.

 From Section 2, we know
that for each value measured at a particular value of $Q^2$,
\beq
\ev{z(Q,m)}=\ev{z(Q,m)}_{pert} \ev{z(m)}_{nonpert}.
\eeq
We find the first term via the Altarelli--Parisi evolution we described in
Section~\ref{1}, and use the measured value (actually a range of possible
values) to determine the nonperturbative factor.

In Tables 2--4, we present the factor $\ev{z(Q,m)}_{pert}$ for
$Q=10.55$ GeV, $m=1.5$ GeV; $Q=91$ GeV, $m=1.5$ GeV; and
$Q=91$ GeV, $m=4.5$ GeV, where we have
taken $\mu_0=2m,m,m/2$ and the values of $\Lambda_5$
shown in the first column.
With these perturbative factors,
we can extract $\ev{z(m)}_{nonpert}$ from  measurements
of the fragmentation function  at CLEO and
ARGUS and LEP (providing the subleading calculation is adequate).

\begin{table}
\begin{center}
\begin{tabular}{||c|c|c|c||}     \hline
$\Lambda_5$ (MeV) & $\mu_0=m_c/2$ & $\mu_0=m_c$ & $\mu_0=2 m_c$
\\ \hline
75  & 0.842 &  0.865 &  0.888   \\  \hline
125 & 0.794 &  0.832 &  0.865   \\  \hline
175 & 0.743 &  0.800 &  0.845   \\  \hline
225 & 0.685 &  0.769 &  0.826   \\  \hline
275 & 0.618 &  0.736 &  0.809   \\  \hline
\end{tabular}
\end{center}
\caption{$\ev{z(Q,m)}_{pert}$ for $Q=10.55$ GeV, $m=1.5$ GeV.}
\end{table}

\begin{figure}[ht]
\epsfxsize=3.6in
\epsfysize=3in
\hbox to \hsize{\hss\epsffile[24 68 574 738]{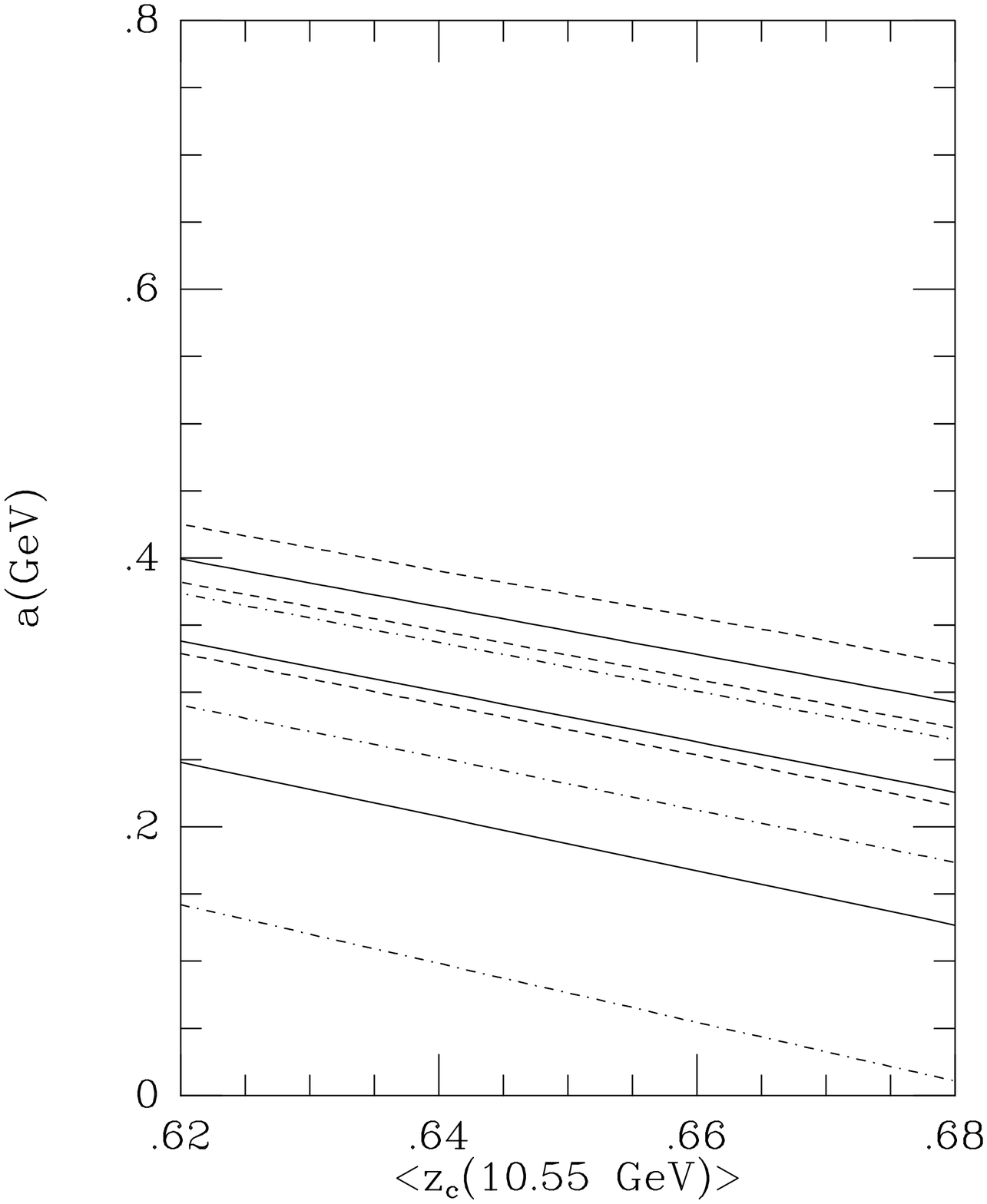}\hskip.5in\hss}
\vspace*{5mm}
\advance\leftskip by 0.36in
\advance\rightskip by 0.36in
{\figfont  Figure 3:
Non-perturbative parameter $a$ (in GeV) as a function of
$\ev{z(Q,m)}$ for $Q=10.55 \gev$ and $m=1.5 \gev$, with
$\Lambda_5=$ 125 MeV (dashed line), 175 MeV (solid),
and 225 MeV (dashed-dotted). For each value of
$\Lambda_5$ we have taken $\mu_0=2m,m,m/2$,
the higher line corresponding to the larger $\mu_0$.}
\end{figure}

The measured values at ARGUS and LEP are the following
\cite{argus}, \cite{aleph}
\bea
\ev{z_c(10.6 \gev)}&=& 0.640 \pm 0.009  \nonumber \\
\ev{z_c(91 \gev)}&=&  0.487 \pm 0.011  \\
\ev{z_b(91 \gev)}&=&  0.714 \pm 0.012   \nonumber
\eea
The numbers from LEP are
still preliminary.
Because the exact numbers are not yet known, and in order that
our analysis can be applied when more exact numbers are measured,
in each case we extract the parameter $a$ over a range of values,
namely
\bea
0.62 \leq &\ev{z_c(10.55 \gev)}& \leq 0.68  \nonumber \\
0.46 \leq &\ev{z_c(91 \gev)}& \leq  0.51  \\
0.69 \leq &\ev{z_b(91 \gev)}& \leq  0.74 \nonumber
\eea

\begin{figure}[ht]
\epsfxsize=3.6in
\epsfysize=3in
\hbox to \hsize{\hss\epsffile[24 68 574 738]{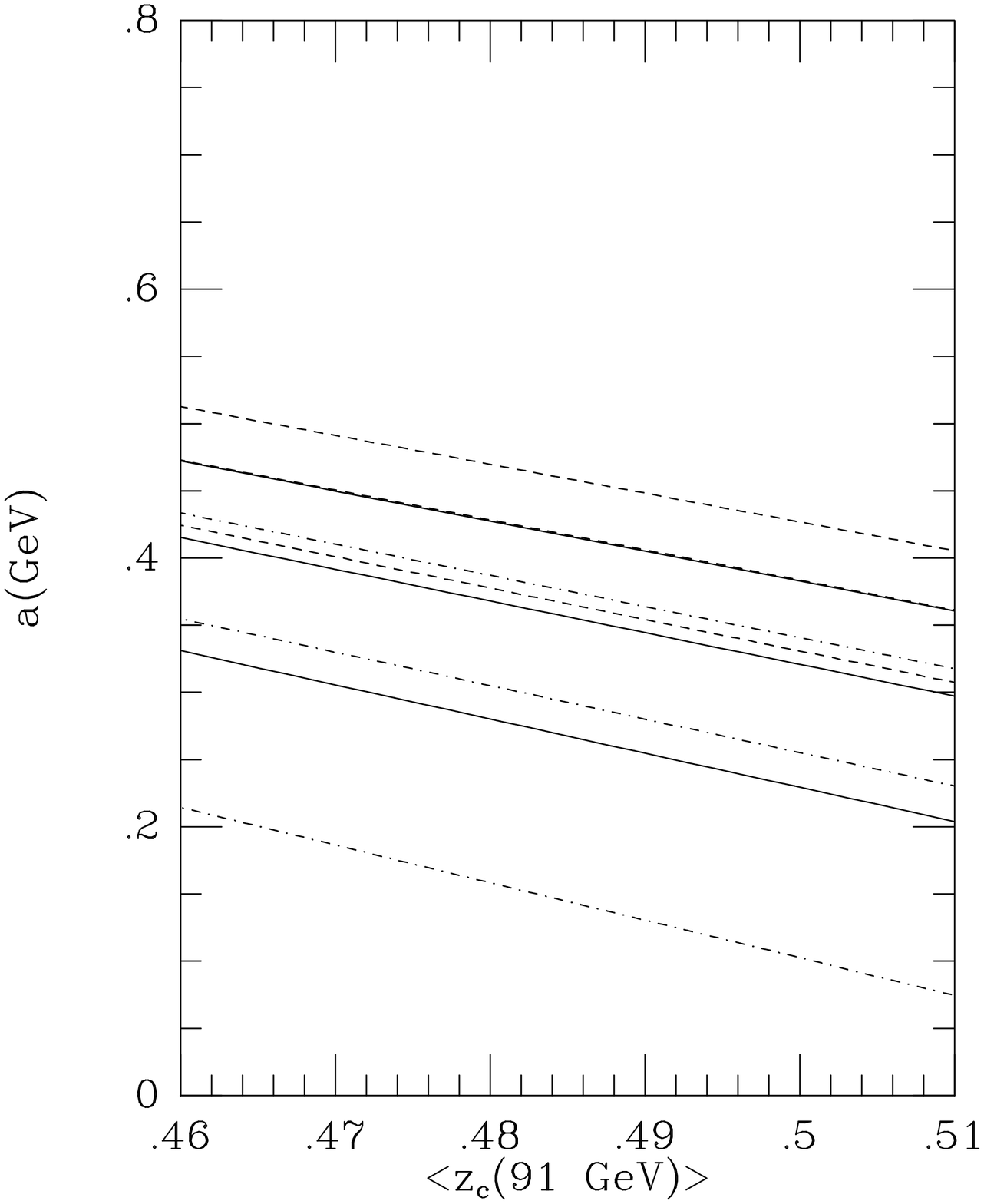}\hskip.5in\hss}
\vspace*{5mm}
\advance\leftskip by 0.36in
\advance\rightskip by 0.36in
{\figfont  Figure 4:
$a$ (in GeV) as a function of
$\ev{z(Q,m)}$ for $Q=91 \gev$ and $m=1.5 \gev$.
The values of $\Lambda_5$ and $\mu_0$
are as in Figure~3.}
\end{figure}

\begin{table}
\begin{center}
\begin{tabular}{||c|c|c|c||}     \hline
$\Lambda_5$ (MeV) & $\mu_0=m_c/2$ & $\mu_0=m_c$ & $\mu_0=2 m_c$
\\ \hline
75  & 0.695 &  0.713 &  0.733   \\  \hline
125 & 0.641 &  0.672 &  0.699   \\  \hline
175 & 0.590 &  0.636 &  0.671   \\  \hline
225 & 0.537 &  0.603 &  0.647   \\  \hline
275 & 0.478 &  0.570 &  0.625   \\  \hline
\end{tabular}
\end{center}
\caption{$\ev{z(Q,m)}_{pert}$ for $Q=91$ GeV, $m=1.5$ GeV.}
\end{table}

Using the perturbative factors from Tables 2-4, in Figures~3-5, we present
(on the vertical axis)
the resulting values of $a$ (in units of GeV) which would be extracted
if the value on the horizontal axis is measured.
We have presented the data graphically, rather than in tabular form
to illustrate the sensitivity to the exact value which
is measured.
The values of $Q^2$ and $m$ are as in the tables.

\begin{table}
\begin{center}
\begin{tabular}{||c|c|c|c||}     \hline
$\Lambda_5$ (MeV) & $\mu_0=m_b/2$ & $\mu_0=m_b$ & $\mu_0=2 m_b$
\\ \hline
75  & 0.829 &  0.836 &  0.846   \\  \hline
125 & 0.807 &  0.816 &  0.828   \\  \hline
175 & 0.789 &  0.800 &  0.814   \\  \hline
225 & 0.772 &  0.786 &  0.802   \\  \hline
275 & 0.757 &  0.773 &  0.791   \\  \hline
\end{tabular}
\end{center}
\caption{$\ev{z(Q,m)}_{pert}$ for $Q=91$ GeV, $m=4.5$ GeV.}
\end{table}

\begin{figure}[ht]
\epsfxsize=3.6in
\epsfysize=3in
\hbox to \hsize{\hss\epsffile[24 68 574 738]{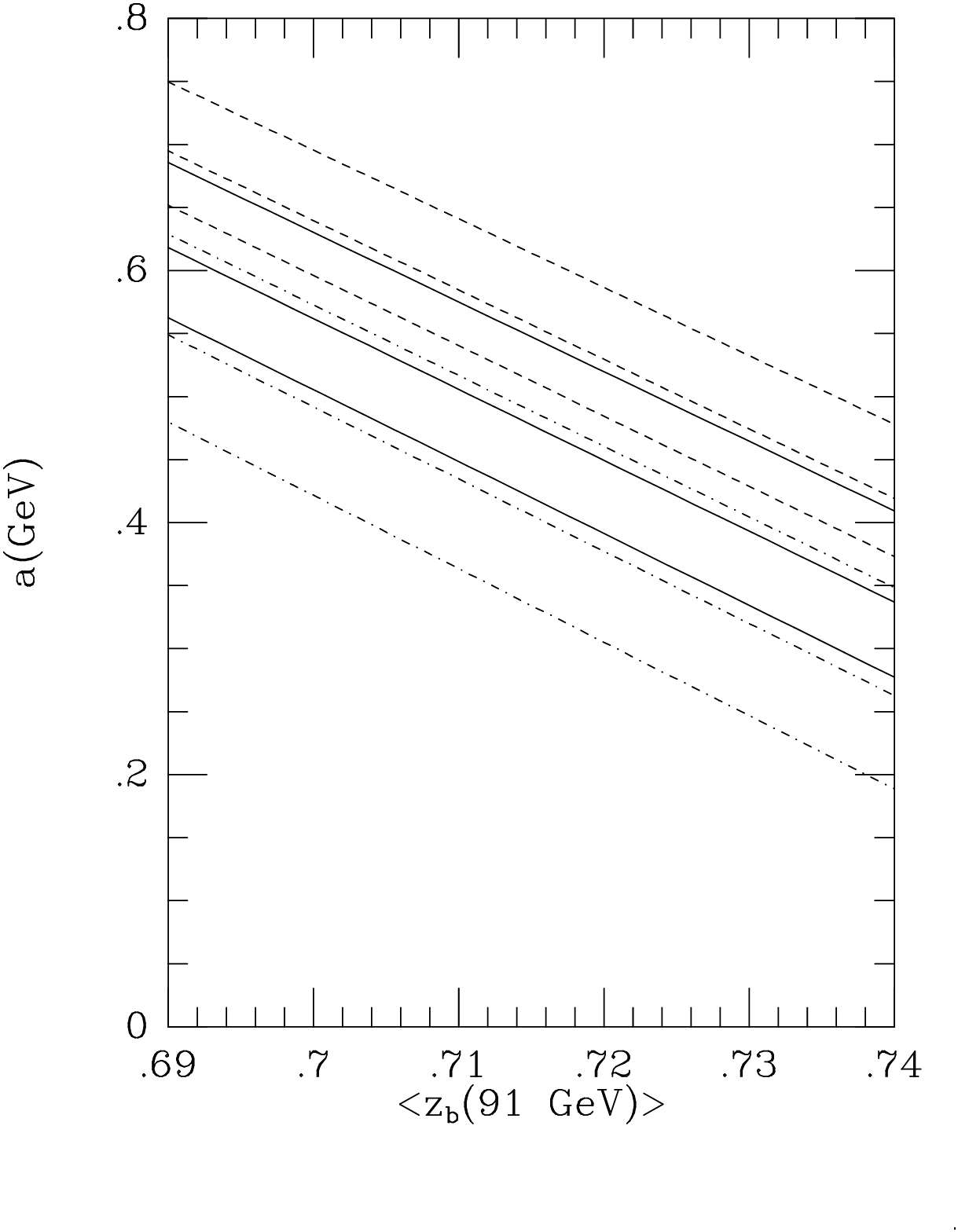}\hskip.5in\hss}
\vspace*{5mm}
\advance\leftskip by 0.36in
\advance\rightskip by 0.36in

{\figfont  Figure 5:
$a$ (in GeV) as a function of
$\ev{z(Q,m)}$ for $Q=91 \gev$ and $m=4.5 \gev$.
The values of $\Lambda_5$ and $\mu_0$
are as in Figure~3.}

\end{figure}

In each figure we give the extracted parameters
for $\Lambda_5=$ 125 MeV (dashed line), 175 MeV (solid line),
and 225 MeV (dotted line) (the current range in the particle
data book \cite{pdb}).
For each value of $\Lambda_5$
we present three lines, corresponding to
$\mu_0=2m,m,m/2$.
The largest values of $a$ in each case correspond to the
largest value of $\mu_0$. This is
readily understood, since there is more QCD scaling for smaller $\mu_0$,
so a smaller $a$ is required to get a particular measured value.

By way of illustration, we give the predictions for $a$
corresponding to $\Lambda_5=$ 175 MeV. First consider
the mean value of the measurement.
In Figure~3, one can see that from the ARGUS measurement
$\ev{z_c(10.6 \gev)} = 0.640$ we predict
$208 \mev \leq a \leq 365 \mev$.
Similarly, from the LEP measurements
$\ev{z_c(91 \gev)} = 0.487$ and
$\ev{z_b(91 \gev)} = 0.714$
in Figures~4-5 we obtain
$264 \mev \leq a \leq 413 \mev$ and
$425 \mev \leq a \leq 552 \mev$, respectively.
Therefore we find
no overlap between the $a$ parameter determined from
the measured mean values $\ev{z_c}$ and $\ev{z_b}$.
For $\Lambda_5=$ 125 MeV there is no overlap either,
and for $\Lambda_5=$ 225 MeV there is overlap
only between the LEP measurements.
However, if we allow for a $2 \sigma$ variation
around the $\ev{z}$ measured in each experiment
there is substantial overlap:
from the ARGUS measurement of $\ev{z_c}$ we predict
$175 \mev \leq a \leq 396 \mev$,
and from $\ev{z_c}$ at LEP
$208 \mev \leq a \leq 461 \mev$, while from $\ev{z_b}$
we obtain
$287 \mev \leq a \leq 686 \mev$.

There are  certain qualitative features of agreement which are
good to note. First, the parameter $a$ is never larger than
800 MeV, and is very likely smaller (especially if the value
is indeed in the overlap region of the $c$ and $b$ quark results).
This is reassuring, as one could not presume to do a heavy quark expansion
for the $c$ quark for a parameter $a$ much larger than this.  Furthermore,
we see there is indeed substantial overlap for each possible value of
$\Lambda_{QCD}$ between the allowed range of $a$ for each experiment.
Remember, according to the heavy quark effective theory, one would
predict the parameter $a$ is the same in each case up to higher order
corrections, of order ${(\Lambda / m)^2}$, which one expects to be
of order 10\%.
It is also clear that the allowed values of $a$ as determined
from the two $\ev{z_c}$ measurements are in very good
agreement.
Both LEP measurements have greater overlap than
the ARGUS measurement has with $\ev{z_b}$, therefore
a measurement from CLEO would be very useful.

\begin{figure}[ht]
\epsfxsize=3.6in
\epsfysize=3in
\hbox to \hsize{\hss\epsffile[24 68 574 738]{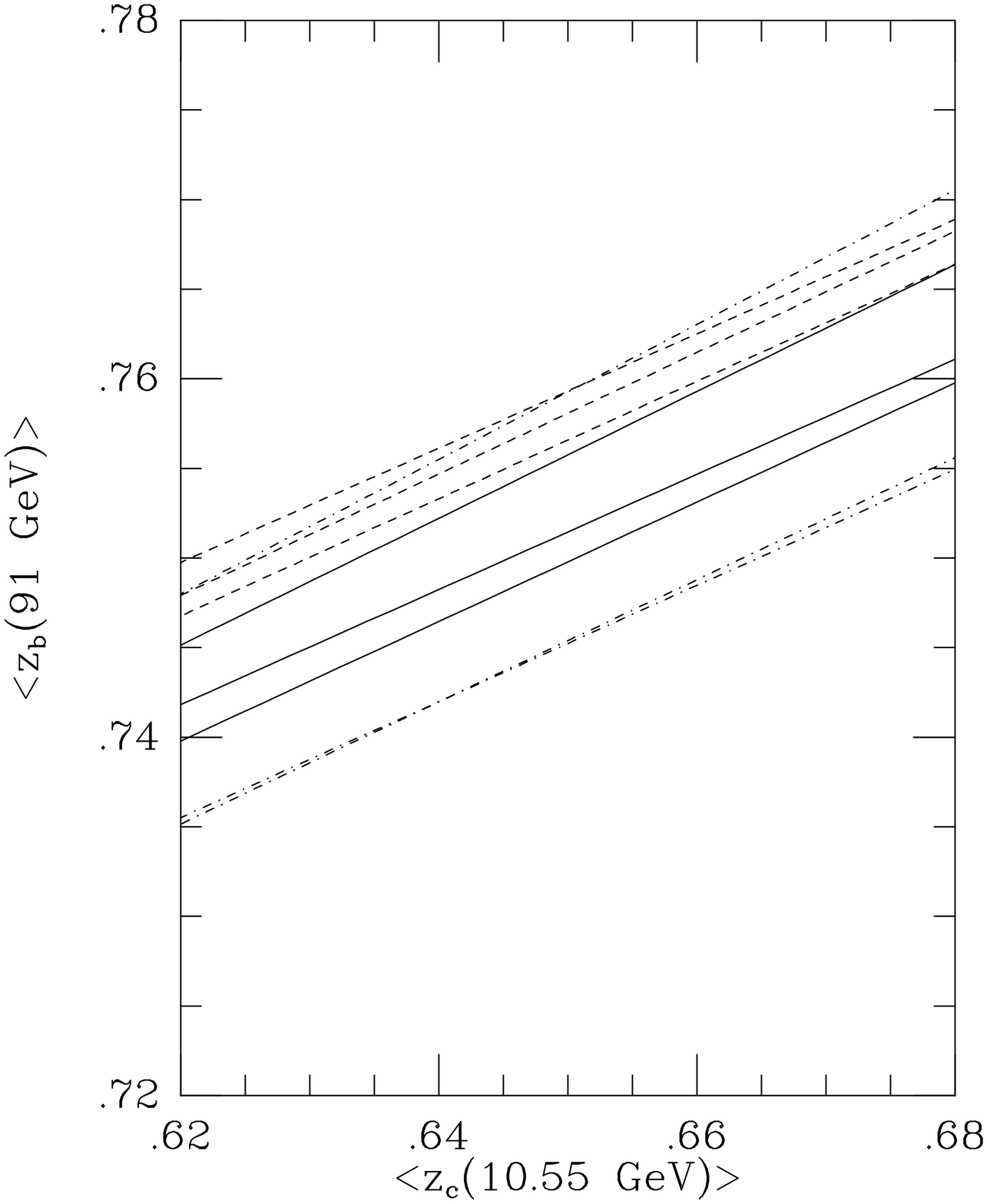}\hskip.5in\hss}
\vspace*{5mm}
\advance\leftskip by 0.36in
\advance\rightskip by 0.36in

{\figfont  Figure 6:
Prediction for $\ev{z_b(91 \gev)}$ as a function of the
measured $\ev{z_c(10.55 \gev)}$,
with $\mu_b/m_b = \mu_c/m_c$.
$\Lambda_5=$ 125 MeV (dashed line), 175 MeV (solid),
and 225 MeV (dashed-dotted).}
\end{figure}

\begin{figure}[ht]
\epsfxsize=3.6in
\epsfysize=3in
\hbox to \hsize{\hss\epsffile[24 68 574 738]{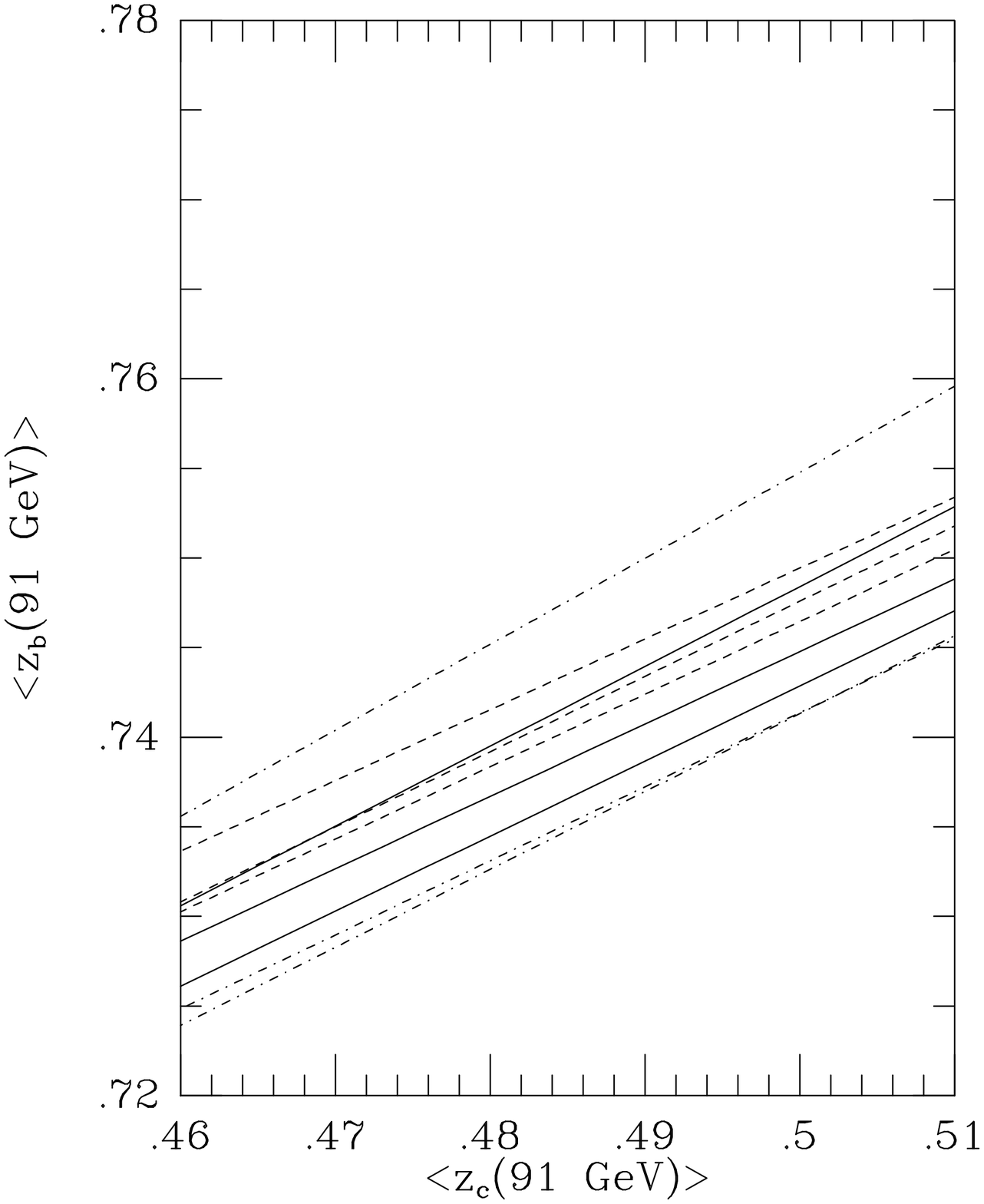}\hskip.5in\hss}
\vspace*{5mm}
\advance\leftskip by 0.36in
\advance\rightskip by 0.36in

{\figfont  Figure 7:
Prediction for $\ev{z_b(91 \gev)}$ as a function of the
measured $\ev{z_c(91 \gev)}$,
with $\mu_b/m_b = \mu_c/m_c$
and the same values of $\Lambda_5$ as in Figure~6.}
\end{figure}

\begin{figure}[ht]
\epsfxsize=3.6in
\epsfysize=3in
\hbox to \hsize{\hss\epsffile[24 68 574 738]{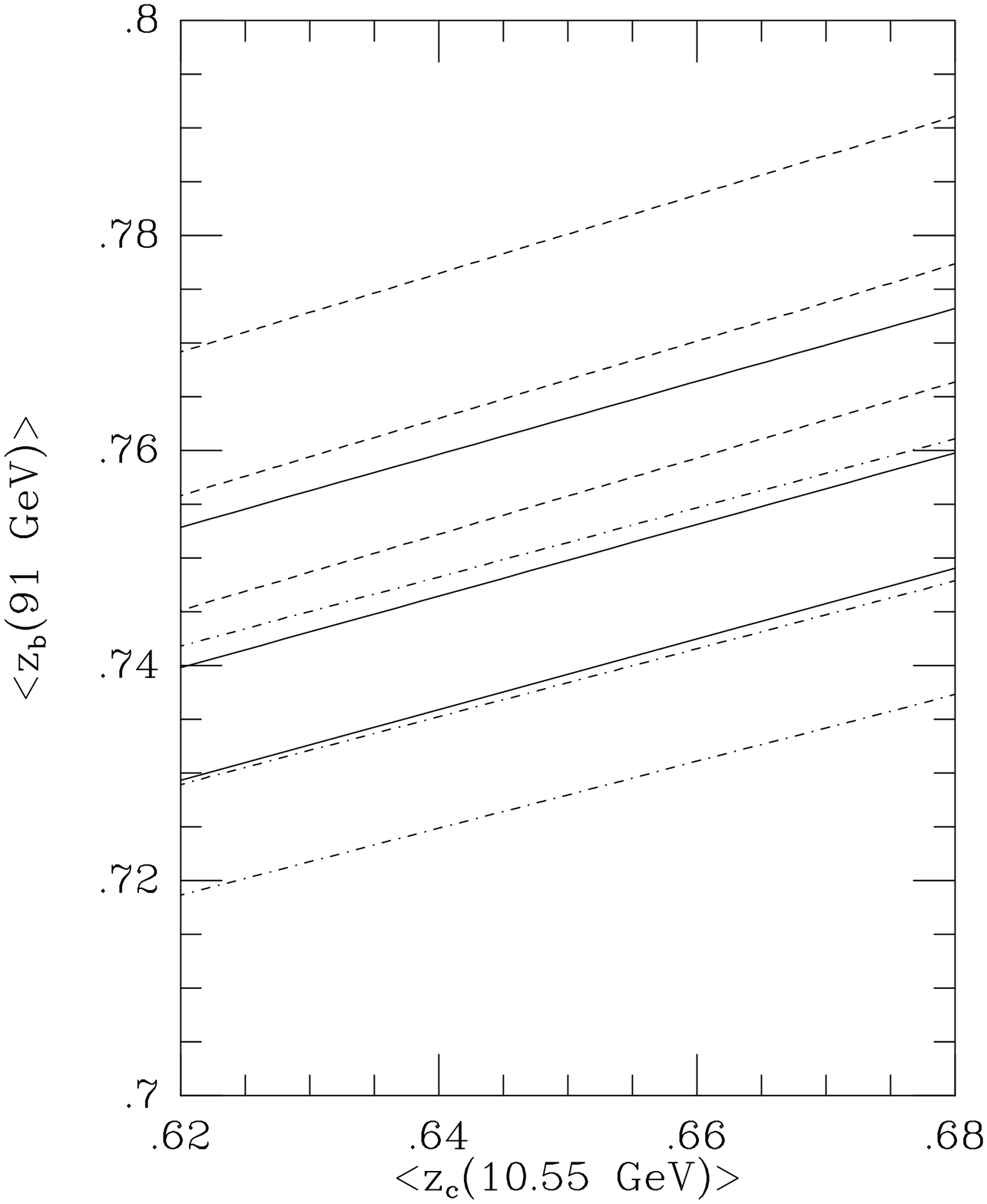}\hskip.5in\hss}
\vspace*{5mm}
\advance\leftskip by 0.36in
\advance\rightskip by 0.36in

{\figfont  Figure 8:
Prediction for $\ev{z_b(91 \gev)}$ as a function of the
measured $\ev{z_c(10.55 \gev)}$, for $\Lambda_5=$ 175 MeV.
For each value of the matching scale
$\mu_c = m_c/2$ (dashed line),
$\mu_c = m_c$ (solid line) and
$\mu_c= 2 m_c$ (dashed-dotted line)
we plot the results for
$\mu_b = m_b/2, m_b, 2 m_b$,
the higher line corresponding to the larger $\mu_b$.}
\end{figure}

\begin{figure}[ht]
\epsfxsize=3.6in
\epsfysize=3in
\hbox to \hsize{\hss\epsffile[24 68 574 738]{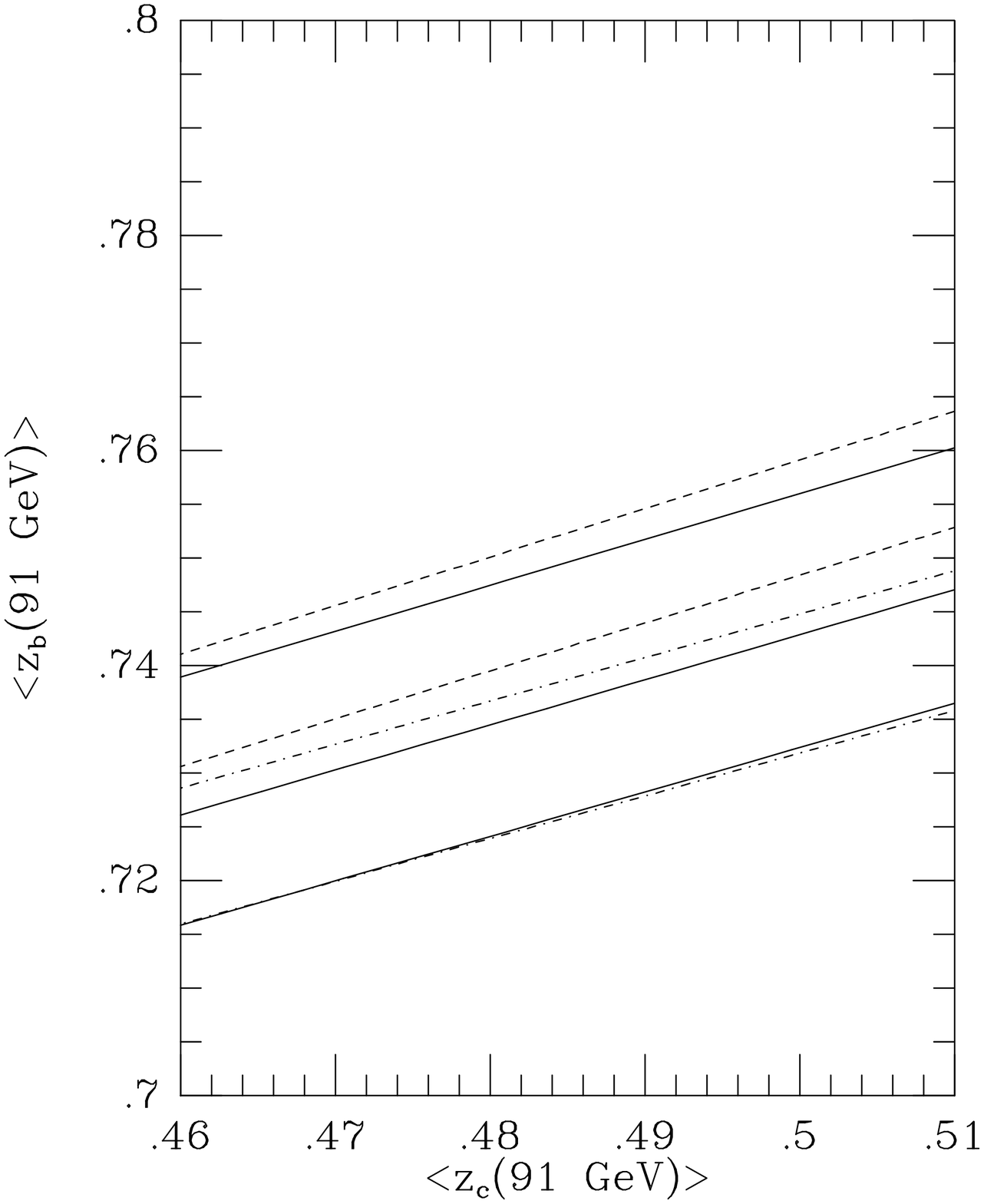}\hskip.5in\hss}
\vspace*{5mm}
\advance\leftskip by 0.36in
\advance\rightskip by 0.36in

{\figfont  Figure 9:
Prediction for $\ev{z_b(91 \gev)}$ as a function of the
measured $\ev{z_c(91 \gev)}$, for $\Lambda_5=$ 175 MeV.
The values of $\mu_c$ and $\mu_b$
are as in Figure~8, although
given that they are somewhat correlated
we omit in this figure the combinations
$\mu_c = m_c/2, \mu_b = 2 m_b$ and
$\mu_c = 2 m_c, \mu_b = m_b/2$.}
\end{figure}

However, the value of the nonperturbative parameter $a$  which is extracted
is clearly strongly  dependent both on the precise  value of $\Lambda_{QCD}$
and the renormalization scale $\mu_0$, indicating the importance
of subsubleading logarithmic corrections.  This is because the
difference between the measured value and the perturbative
contribution is much more strongly $\mu_0$--dependent (defined
as the fraction of the value of the quantity) than the perturbative
contribution itself. So for example, if $\ev{z_b}$ measured
at LEP is $0.71$, the parameter $a$ for $\Lambda_5=125$ MeV,
can vary between about 537 MeV and 637 MeV,
even though the perturbative
factor varies only between $0.81$ and $0.83$. In order to determine
an effect of size $\Lambda_{QCD}/m$, the perturbative contribution
needs to be known at accuracy more precise than this. Clearly,
the subleading calculation is not adequate for  a
precise extraction of $a$. This problem is not endemic
to the particular test of the
heavy quark theory we have done. It seems likely that in
order to determine $\Lambda/m$ corrections, one either
must take ratios in which the QCD corrections cancel, or
do an even higher order calculation.

 However, the parameter $\ev{z}_{nonpert}$
is more stable against QCD uncertainties. This is actually the
quantity upon which the measured value depends.  When
we  relate two different measured values of $\ev{z}$,
the variation with $\mu_0$ and $\Lambda_{QCD}$ might not be so large.
That is, we have not at all correlated the choice of $\mu_0/m$
for the $b$ and $c$ quarks.

In Figures 6-7, for the assumed values of
$\ev{z_c(Q)}$ which are given on the horizontal axis, we determine
the nonperturbative parameter $a$. We then use this to predict
the value of $\ev{z_b(91 \gev)}$ which should be measured for the
$b$ quark (this was the procedure followed in ref. \cite{cn} but for
the fragmentation function itself).
There are two graphs, corresponding to $Q=10.55$ GeV and
$Q=91$ GeV.
In each case, we have allowed for $\Lambda_5=125,175,225$ MeV.
For each of these, we have allowed for $\mu_0=2m,m,m/2$.

Notice that the current measurement looks fairly borderline
in that $\langle z_b \rangle$ is lower than the favored
range. However, this procedure, which is essentially that
of \cite{cn}, probably gives too narrow a range of predictions,
since it is probably
not justified to always choose exactly
the same value of $\mu_0/m$.  This procedure underestimates
the $\mu$ variation, because the
values of $a$ one extracts are highly correlated with respect to
the choice of $\mu_0/m$.  So for example,
we get a similar prediction if we take $\mu_0=m_c$ to extract $a$
and use $\mu_0=m_b$ to determine $\ev{z_b}$ as to if we had
used $\mu_0=2m_c$ and $2m_b$ respectively.

The optimal choice of $\mu_0$ would correspond to that for which higher order
terms are negligible, so the subleading order answer, evaluated at
the renormalization point $\mu_*$, is the correct answer.  One
can then ask the question whether $r=\mu_*/m$ depends on the mass, $m$.
It is straightforward to see that it is roughly
independent of $m$ if one
is performing perturbation theory in $\alpha$. However, for a leading
log calculation, $r$ will be formally mass independent only so long
as the ratio $Q/m_1$ is much greater than $m_1/m_2$, where $Q$ is
the center of mass energy, $m_1$ will be the $b$ quark mass, and $m_2$
will be that of the $c$ quark (assuming common $Q$). In reality,  one probably
requires
that  the scaling between
$Q$ and $m_1$ is much greater than that between $m_1$ and $m_2$.

We therefore conservatively allow for independent $\mu_0$ variation
in the $b$ and $c$ quark calculations. In the case where there are
two values of $Q$ involved, we have allowed the full $\mu_0$ variation
of each. In the case where $Q=M_Z$ for both the $b$ and $c$ quarks,
we have always taken both $\mu_0$ greater than $m$ or both less than $m$,
given that the value of $\mu_*$ is somewhat correlated. This
is probably a fairly conservative range. The  range might be less.

In Figures 8 and 9, we  give the allowed
range of possible predictions for $\ev{z_b}$ when $\Lambda_5=175$ MeV,
given a measurement
of $\ev{z_c}$ at $Q=10.55 \gev$ and $Q=91 \gev$ respectively.
The corresponding curves for $\Lambda_5=125\ \mev$ are slightly
more predictive and for $\Lambda_5= 225\ \mev$
are somewhat less predictive.
The allowed range of predictions is greater than in Figure 6 and 7,
but still somewhat predictive.
For example, if we vary $\ev{z_c}$ within $2 \sigma$ around the
measured value
$\ev{z_c(10.55 \gev)} = 0.640$ we obtain
$0.719 \leq  \ev{z_b(91 \gev)} \leq 0.782$,
and similarly for $\ev{z_c(91 \gev)} = 0.487$ we get
$0.718 \leq  \ev{z_b(91 \gev)} \leq 0.763$
within $2 \sigma$.

Figures 8 and 9 are our main results. They tell us
which measurements would be consistent with the heavy
quark expansion, in light of  the perturbative QCD uncertainties.
We see that the range is sufficiently narrow
that tests of the heavy quark expansion are possible. A
measurement outside this range would be an indication of a $(\Lambda/m)^2$
contribution to $\langle z \rangle_{nonpert}$ (or less likely,
an indication of nonstandard model physics in low energy
QCD scaling).  We see
that the preliminary numbers do not yet require such higher order terms.
Reducing the error may or may not leave us inside the range
of predictions given in Figures 8-9.

\section{Cutoff Moments}
\label{3}

In general, when extracting the moments experimentally, the
data does not go down to arbitrarily low $z$. It is therefore
necessary to extrapolate the data into this region. This
however introduces model dependence. The model to which one
fits will have some effect on the number which is extracted.
An uncertainty due to this modeling error can be assigned by
attempting to fit to a couple of different functions; however
this does not necessarily represent the true error.

In this section, we argue that it might be better to work with truncated
moments. By this we mean one can define a moment as the
integral from $z_0$ to 1, rather than from 0 to 1. Ideally,
one can choose
 $z_0$ as the
experimental cut, so long as $z_0$ is not too large.

It is however not standard to do this, because unless one integrates to zero,
the equation for the evolution of the moments does not factorize.
In particular, when one defines
\beq
D_{z_0}^N=\int_{z_0}^1 dz z^{N-1} \hat{f}(z),
\eeq
and uses
\beq
{d \f(z) \over d \log{\mu^2}}
=\int_z^1 P(z/y) \f(y) {dy \over y} \ ,
\eeq
one finds
\beq
{d D^N_{z_0} \over d \log \mu^2}=
\int_{z_0}^1 dy y^{N-1} \f(y) \int_{z_0/y}^1 dz
z^{N-1} P(z)
\eeq
In the case $z_0=0$ we see this equation factorizes into the product
of moments. However, because of the $y$ dependence of the endpoint
on the $z$ integration, this is not the case for general $z_0$.
The problem here is to evaluate the right hand side, you would
actually need to know $\f(y)$, which is precisely what we wish to avoid.
What we would like is an expression solely in terms of moments, so
we can follow the same procedure described in section~\ref{1}.

In this section we will focus again on the second moment
and we will show that if we make the approximation
that the evolution equation {\it does} factorize, by taking the lower
limit on the $z$ integration to be $z_0$, the error we make is
actually only of order $z_0^2$ which is generally rather small.

So we consider the difference
\beq
\int_{z_0}^1 dz z P(z)-\int_{z_0/y}^1 dz z P(z)
=\int_{z_0}^{z_0/y} dz z P(z)
\eeq
between the function which truly appears in the evolution equation and
the approximation one obtains by taking the lower endpoint of the integral
to be $z_0$.
We consider only the leading log anomalous dimension, as the error
from the subleading piece is suppressed by $\alpha_{QCD} $ and should
be a small correction compared to the
error from the leading term.

As discussed in section~\ref{1}, it
is sufficient to consider only nonsinglet evolution which involves the function
\beq
P(z)\propto\left({1+z^2 \over 1-z}\right)_+
\eeq
It should be kept in mind that the ``+" keeps the function
properly normalized. However this should be true independent of
the arbitrary choice of  $z_0$.  We therefore take the coefficient of the
$\delta$ function which needs to be subtracted to be determined from
integrating over all $z$ from 0 to 1. (The
fact that this assumption leads to a smaller error is good confirmation
of the expectation that this will better approximate the true function).
So the error in our approximation is determined from the following difference
\beq
\int_{z_0}^{z_0/y} dz z {1+z^2 \over 1-z}
\eeq
which is
\beq
{3 z_0^2 \over 2}(({1\over y})^2-1)+\ldots
\eeq
where the $\ldots$ represents higher order terms in $z_0$
and we have expanded the logarithm which is obtained upon integration.
 The
error is only of order $z_0^2$. Really, it is of order $(z_0/y)^2$.
However, so long as the distribution is weighted towards unity,
as it is for a the fragmentation function of a heavy
quark evaluated at a sufficiently low momentum scale,
our statement is approximately correct.

To check that we do not make too large
an error when applying this approximation
to a heavy quark fragmentation function,
 we explicitly evaluated the error for different
assumed functional forms which were claimed to fit data reasonably
well in the past. Notice that we are using the function only
to estimate the error, but not to fit the data. Furthermore,
the error estimate only involves those values of $z$ which
are measured--one never needs the values of the function
below $z_0$. So ultimately, it is possible to estimate
the error based on a function which fits the data, without
extrapolation.

The test functions we used
were a Peterson function with a large range of $\epsilon$
and also a function of the form from ref. \cite{cn}, namely
of the form $z^\alpha (1-z)^\beta$.  The
resulting differences (in percent)
between the exact and approximate evolution
of the second moment are given in Tables 5 ($c$ quark)
and 6 ($b$ quark) for a range of cutoff values, $z_0$.
We see that in no case does the error exceed 10\%
for the $c$ quark and 2\% for the $b$ quark, and
in most of the cases it is considerably smaller.
These particular numbers correspond to $Q=M_Z$,
and we have taken $\Lambda_5 = 200 \mev$
and $\mu_0 = m$, but varying them in the ranges we have allowed for
in section~2 does not change our conclusions.
It is important
that the error is calculated entirely based on the region above
$z_0$, so no assumption is made about extrapolating to
the unmeasured region. It can evaluated in a fairly model
independent manner.

It is clear that the error is always
within the regime of accuracy of our predictions (ie
smaller than the uncertainty from higher order QCD effects). Therefore,
it is as useful to determine the cutoff moments as it is to  determine
the true ones, and probably has a smaller systematic error. We suggest the
data be presented in the future in this way when there is a cut on $z_0$ (or
as a function of cut values for comparison with other experiments).  As we
have seen, the error when using our procedure to extract meaningful numbers
should be quite small, of order a few percent.

\begin{table}
\begin{center}
\begin{tabular}{||c|c|c|c|c||}     \hline
$z_0$            &  0.05  & 0.1 & 0.15 &  0.2   \\ \hline
$\alpha = 0.35$, \ $\beta=3.05$  &   0.18  & 0.9 & 2.0 & 3.6  \\ \hline
$\epsilon_c=0.14$  & 0.4  &  1.7  & 3.7  & 6.6 \\ \hline
$\epsilon_c=0.44$  & 0.7  &  2.8  & 6.0  & 10.  \\ \hline
\end{tabular}
\end{center}
\caption{Relative difference (in $\%$) between the exact and
approximate evolution of the second moment for the $c$ quark.}
\end{table}

\begin{table}
\begin{center}
\begin{tabular}{||c|c|c|c|c||}     \hline
$z_0$            &  0.05  & 0.1 & 0.15 &  0.2   \\ \hline
$\alpha=0.595$, \ $\beta=18.67$  &     & 0.15 & 0.3   &  0.6     \\ \hline
$\epsilon_b=0.016$  & 0.05  &  0.18  & 0.7  & 1.3 \\ \hline
$\epsilon_b=0.049$  & 0.1  & 0.4  & 1.0 & 2.0  \\ \hline
\end{tabular}
\end{center}
\caption{Relative difference (in $\%$) between the exact and
approximate evolution of the second moment for the $b$ quark.}
\end{table}

\section{Conclusions}

In this paper, we have used the measured values of
the heavy quark fragmentation second moments
to test the heavy quark effective theory.
By relating $c$ and $b$ quark measurements we have
extracted a mass suppressed heavy quark parameter.
We have also related measurements of $c$ quark fragmentation
using purely perturbative QCD. The graphs of Figures 1, 8, and 9
are our main results.

 From the purely perturbative analysis,
we have seen that the measurements of $c$ quark fragmentation
at different center of mass energies are consistent
with QCD predictions, and tend to favor large
values of $\Lambda_{QCD}$. This part of the
analysis does not involve heavy quark symmetry and could prove
an alternative way of determining the strong coupling.

We have found the value of the nonperturbative
parameter $a$ as measured with the $c$
quark lies between
$79 {\rm MeV}$ and $422 {\rm MeV}$ (assuming
it is in the overlap of the two measurements) and in the
range $199 {\rm MeV}$ to $750 {\rm MeV}$ for the $b$ quark.
A better determination of $\Lambda_{QCD}$ would narrow
the allowed range of $a$ sizably.
For instance, we have obtained that if $\Lambda_5=$ 175 MeV
$a$ lies between
$208 {\rm MeV}$ and $396 {\rm MeV}$ for the $c$ quark and in the
range $287 {\rm MeV}$ to $686 {\rm MeV}$ for the $b$ quark.
Although the range is substantial, it is of interest to have
any measurement at the level of $\Lambda/m$.  Furthermore,
it is encouraging that these values are reasonably small
and overlap. With better measurement, one might hope to
search for deviations at the level of $(\Lambda/m)^2$. However,
given the large perturbative QCD uncertainty, it is not yet clear
whether this will be possible.

Finally, we have suggested to use ``cutoff'' moments, which do
not involve extrapolation of the experimental data to low $z$,
and therefore will probably have smaller systematic error.
We have shown that although factorization is not exact,
it is a good approximation within the regime of accuracy of
our calculation.

There are several measurements which it would be nice to
see done or improved. One would want CLEO to measure
the ratio of $D$ to $D^*$ production to test for the importance
of higher twist effects. One
would also want to take advantage of the many $c$
quark states at CLEO to get a better low energy
measurement of the fragmentation function.
 A measurement of the production of $c c \overline{c}
\overline{c}$ at LEP would be a useful confirmation of the claim that
multi $c$ quark production will not distort our predictions.
A good exclusive measurement on $D^*$'s at LEP would be very
useful when using fragmentation functions to extract $\Lambda_{QCD}$.

It would certainly be advantageous to improve the measurements
of heavy quark fragmentation functions.
  Heavy quark fragmentation
could prove to be an interesting test of the heavy
quark mass expansion and a supplement to existing measurements
of $\alpha_{QCD}$.

\section*{Acknowledgements} We would like to thank Leo Bellantoni,
Giovani Bonvicini, David Lambert, and Vivek Sharma
for providing us with information on existing measurements.
We would also like to thank Adam Falk, Ian Hinchliffe, Bob Jaffe,
Costas Kounnas, Mike Luke,
Misha Shifman, and Mark Wise for useful conversations. We are
especially grateful to Paolo Nason for his assistance.
We thank the CERN Theory Division and the Institute for Theoretical
Physics in Santa Barbara for their hospitality.

\end{document}